\documentclass{amsart}%
\usepackage{graphicx}
\usepackage{amscd}
\usepackage{amsmath}
\usepackage{amsfonts}
\usepackage{amssymb}%
\newtheorem{theorem}{Theorem}
\theoremstyle{plain}

\newtheorem{corollary}{Corollary}

\newtheorem{definition}{Definition}
\newtheorem{example}{Example}

\newtheorem{lemma}{Lemma}

\newtheorem{proposition}{Proposition}
\newtheorem{remark}{Remark}

\numberwithin{equation}{section}

\begin{document}
\title{A hyper-geometric approach to the BMV-conjecture}

\begin{abstract}
We provide a representation of the (signed) BMV-measure by stochastic means
and prove positivity of the respective measures in dimension $d=3$ in several
non-trivial cases by combinatorial methods.

\end{abstract}
\author{Michael Drmota, Walter Schachermayer, Josef Teichmann}
\address{Institute of Discrete Mathematics and Geometry and Institute of Mathematical
Methods in Economics, Vienna University of Technology, Wiedner Hauptstrasse
8-10, A-1040 Vienna, Austria}
\email{michael.drmota@tuwien.ac.at, wschach@fam.tuwien.ac.at, \linebreak josef.teichmann@fam.tuwien.ac.at}
\thanks{The second author was supported by the Wittgenstein prize program Z-36. Much
of this work has been done, while he was visiting the university Paris 9. He
thanks I. Ekeland and E. Jouini for their hospitality. The third author had
the opportunity to spend one month at the university Paris 7 and three weeks
at CREST. He thanks in particular for the kind hospitality of Laurence
Carassus and Nizar Touzi. The initial motivation for our research on the
problem happened during conversations with Peter Michor and Martin Feldbacher.
The authors are indebted to M. Fannes and R. Werner who kindly offered us a
preprint on known results concerning the BMV-conjecture.}
\maketitle

\baselineskip 0.25in

\section{Introduction and Results}

We aim to provide a representation of the (signed) measure related to the
Bessis-Moussa-Villani conjecture (in the sequel BMV) by stochastic methods and
calculate non-trivial cases in dimension $3$ by hyper-geometric methods.

\begin{definition}
Let $d\geq1$ be fixed. Let $A$, $B$ be complex, hermitian $d\times d$ matrices
and $B\geq0$, then we denote%
\[
\phi^{A,B}(z):=\operatorname*{tr}(\exp(A-zB))
\]
for $z\in\mathbb{C}$.

The Bessis-Moussa-Villani conjecture (open since 1975, see \cite{BesMouVil:75}%
) asserts that the function $\phi^{A,B}$ is completely monotone, i.e.
$\phi^{A,B}$ is the Laplace transform of a \textbf{positive} measure
$\mu^{A,B}$ supported by $[0,\infty\lbrack$,%
\[
\operatorname*{tr}(\exp(A-zB))=\int_{0}^{\infty}\exp(-zx)\mu^{A,B}(dx).
\]
Since the function $\phi^{A,B}$ is always Laplace transform of a possibly
signed measure on $[0,\infty\lbrack$, we shall always denote this signed
measure by $\mu^{A,B}$.\label{bmvdef}
\end{definition}

The BMV conjecture is closely related to convergence assertions on
perturbation series in quantum mechanics and there is a substantial literature
on it (recently \cite{LieSei:03} has been published, where several further
references can be found). We quote from \cite{Wer03}: \textquotedblright The
BMV conjecture would entail a number of interesting inequalities not just for
quantum partition functions, but also for their derivatives (a badly needed
tool). Despite a lot of work, some by prominent mathematical physicists, only
some simple cases have been decided. So far all results, including fairly
extensive numerical experiments, are in agreement with the
conjecture\textquotedblright. As an example of recent progress on positive
results we mention that the BMV-conjecture was shown to hold true in an
average sense in \cite{FanPet:01}. Originally the BMV-conjecture was
formulated more generally, namely, that $z\mapsto\left\langle e,\exp
(A-zB)e\right\rangle $ is completely monotone for each eigenvector $e$ of $B$.
This first conjecture was seen to be wrong immediately (see the end of the
article \cite{BesMouVil:75}). In Appendix 2 we provide a simple
counter-example for the sake of completeness.

At first sight the following relations hold true:

\begin{proposition}
Let $A,B$ be hermitian $d\times d$ matrices, $B\geq0$, then $\phi^{A,B}%
(z)\geq0$ for $z\geq0$ and%
\begin{align*}
\frac{d}{dz}\phi^{A,B}(z)  &  =-\operatorname*{tr}(\exp(A-zB)B)\\
\frac{d^{2}}{dz^{2}}\phi^{A,B}(z)  &  =\operatorname*{tr}\left(  \int_{0}%
^{1}\exp(-s(A-zB))B\exp(s(A-zB))Bds\exp(A-zB)\right)
\end{align*}
for $z\geq0$. Hence $-\frac{d}{dz}\phi^{A,B}(z)\geq0$ and $\frac{d^{2}}%
{dz^{2}}\phi^{A,B}(z)\geq0$ for $z\geq0$.
\end{proposition}

\begin{proof}
The first assertion follows from the fact that the eigenvalues of $\exp(A-zB)$
are non-neagtive and the second from the derivative of the function $\exp$ off
$0$ (see for instance \cite{KriMic:97}, Theorem 38.2),%
\[
\frac{d}{dz}\exp(A-zB)=-\exp(A-zB)\int_{0}^{1}\exp(-s(A-zB))B\exp(s(A-zB))ds.
\]
Hence%
\begin{align*}
\frac{d}{dz}\phi^{A,B}(z)  &  =-\operatorname*{tr}\left(  \exp(A-zB)\int
_{0}^{1}\exp(-s(A-zB))B\exp(s(A-zB))ds\right) \\
&  =-\int_{0}^{1}\operatorname*{tr}\left(  \exp(A-zB)\exp(-s(A-zB))B\exp
(s(A-zB))\right)  ds\\
&  =-\int_{0}^{1}\operatorname*{tr}(\exp(A-zB)B)ds\\
&  =-\operatorname*{tr}(\exp(A-zB)B).
\end{align*}
The second formula follows by a similar reasoning. We conclude the
inequalities in a \textquotedblright moving frame\textquotedblright%
\ associated to the eigenbasis of $\exp s(A-zB)$.
\end{proof}

Bernstein's Theorem (see for instance \cite{Fel:66}) tells that a smooth
function $\phi:\mathbb{R}_{\geq0}\rightarrow\mathbb{R}$ is the
Laplace-transform of a non-negative measure $\mu$ on $\mathbb{R}_{\geq0}$ if
and only if $(-1)^{n}\phi^{(n)}(z)\geq0$ for $z\geq0$. For the BMV-function
$\phi^{A,B}$ we know by the previous Lemma at least, that the Bernstein
condition holds for $n=0,1,2$. For dimension $d\geq3$ the case $n=3$ is
unknown in general. Having Bernstein's Theorem in mind, we see that the
validity of the BMV-conjecture is equivalent to a sequence of interesting
trace inequalities for hermitian matrices.

The following simple transformation properties are immediately proved.

\begin{enumerate}
\item Given a unitary matrix $U$ in dimension $d$, then $\mu^{UA\overline
U^{T},UB\overline U^{T}}=\mu^{A,B}$. This is due to the unitary invariance of
the trace functional.

\item Let $I_{d}$ denote the identity matrix in dimension $d$. Then
$\mu^{A+\lambda_{1}I_{d},B}=\exp(\lambda_{1})\mu^{A,B}$ for all real
$\lambda_{1}$, since the identity matrix commutes with $A,B$.

\item $\mu^{A,B+\lambda_{2}I_{d}}=\mu^{A,B}(.+\lambda_{2})$ for $\lambda
_{2}\geq-b_{\min}$, where $b_{\min}$ denotes the minimal eigenvalue of $B$,
since a translation of $B$ by $\lambda_{2}I_{d}$ corresponds to a translation
of the measure by $\lambda_{2}$.
\end{enumerate}

Furthermore the following cases are known, where the BMV-conjecture holds true.

\begin{enumerate}
\item If $A$ and $B$ commute, the BMV-conjecture holds true.

\item If $d=1,2$, the BMV-conjecture holds true.

\item If $B$ has at most two different eigenvalues, the BMV-conjecture holds true.

\item Let $B$ be a diagonal matrix. If the off-diagonal elements of $A$ are
non-negative, the Dyson expansion (see Section 2 for a stochastic proof)
yields that the BMV-conjecture holds true.
\end{enumerate}

In view of all these well-known facts (for more investigations in these
directions see \cite{FanWer:00}), the first non-trivial case, which appears in
lowest non-trivial dimension, is the following. Take $d=3$, $B=diag(b_{1}%
,b_{2},b_{3})$ a diagonal matrix and $A=(a_{ij})$, and assume $a_{12}%
a_{13}a_{23}<0$. In this article we give -- by hyper-geometric methods -- a
partial positive answer in this case.

We first provide a Feynman-Kac-type construction -- by stochastic means -- of
the Dyson series. Since Feynman-Kac-type Theorems are often used to prove that
a function is a Laplace transform, we were motivated to construct an
appropriate Markov process for the non-stochastic matrix $A$ in order to prove
a Feynman-Kac representation for $\phi^{A,B}$. From this representation we can
deduce the Dyson expansion and we are able to deduce a \textquotedblright
semantics\textquotedblright\ of the problem, however, we were not able to
conclude the result directly. The proof of the Feynman-Kac Theorem can be
found in the Appendix 1, its application in order to prove the Dyson expansion
in Section 2, see Theorem \ref{dyson}.

In Section 3 we concentrate on the $3$-dimensional case, where we meet an
important combinatorial simplification, see \ref{l-k-relation}, which then
leads to a summation problem in the theorie of hyper-geometric series. Finally
we are able to prove the following result.

\begin{theorem}
\label{TH}Given a real, symmetric $3\times3$ matrix $A=(a_{ij})$ and a
diagonal matrix $B=diag(b_{1},b_{2},b_{3})$ with diagonal elements $0\leq
b_{1}<b_{3}<b_{2}$. We assume that the following two conditions hold true:

\begin{enumerate}
\item $\frac{|a_{12}|}{\sqrt{b_{2}-b_{1}}}\geq\frac{|a_{13}|}{\sqrt
{b_{3}-b_{1}}}$ and $\frac{|a_{12}|}{\sqrt{b_{2}-b_{1}}}\geq\frac{|a_{23}%
|}{\sqrt{b_{2}-b_{3}}}$.

\item $a_{11}(b_{2}-b_{3})+a_{22}(b_{3}-b_{1})+a_{33}(b_{1}-b_{2})\geq0$.
\end{enumerate}

Then the function $\phi^{A,B}(z):=\operatorname*{tr}(\exp(A-zB))$ is
completely monotone and the BMV-conjecture holds. Furthermore the
BMV-conjecture holds (trivially) if two of the three eigenvalues $b_{1}%
,b_{2},b_{3}$ agree or $a_{12}a_{13}a_{23}\geq0$.
\end{theorem}

\begin{remark}
The unusual order $b_{1}<b_{3}<b_{2}$ is due to the structure of our proof,
see Section 3. Later we shall assume $b_{1}=0$, which is possible without
restriction of generality as we have noted above under transformation property
(3) on page 3.
\end{remark}

\begin{remark}
The two conditions in (1) are related to positivity on the intervals
$]0,b_{3}[$ and $]b_{3},b_{2}[$, respectively (in this order). The second
condition is a linear functional on the diagonal values of $A$ and appears to
be the same on both intervals.
\end{remark}

\begin{remark}
The proof of Theorem \ref{TH} will be given in Section 5. We note, however,
that the assertions of the last sentence can be proved immediately (the
argument shows all trivial cases).
\end{remark}

\begin{proof}
Assume that $a_{12}a_{13}a_{23}\geq0$, then we can make a change of
coordinates such that $a_{ij}\geq0$, for $i\neq j$ by multiplying two
coordinates by $-1$. If $a_{ij}\geq0$ holds for $i\neq j$, then by Theorem
\ref{dyson} the measure $\mu^{A,B}$ is a sum of non-negative measures, hence
non-negative. If $b_{2}=b_{3}$, then $B$ has a $2$-dimensional eigenspace,
where we can rotate without changing $B$, consequently we can find an
orthogonal matrix $U$ such that $(U^{T}AU)_{23}=0$ and $U^{T}BU=B$. The trace
is invariant under rotations, so%
\[
\phi^{A,B}(z)=\operatorname*{tr}(\exp(A-zB))=\operatorname*{tr}(\exp
(U^{T}AU-zU^{T}BU)),
\]
hence we find ourselves in the first trivial case.
\end{proof}

\section{Representation of $\mu^{A,B}$}

In this section we fix $d\geq2$ and a $d\times d$ hermitian matrix $A$. We
shall construct a Markov process $(Y_{t}^{(\zeta,i)})_{0\leq t\leq1}%
:=(Z_{t}^{(\zeta,i)},X_{t}^{i})_{0\leq t\leq1}$ for $(\zeta,i)\in S$, which
leads via the Feynman-Kac formula (see Theorem \ref{fk2} in the Appendix 1),
to a representation of the BMV-measure $\mu^{A,B}$. The state space of this
Markov process is $S:=\mathbb{C}\times\{1,\dots,d\}$. We shall assume a
filtered probability space $(\Omega,\mathcal{F},(\mathcal{F}_{t})_{0\leq
t\leq1},P)$, which allows for the subsequent constructions. Later we shall
identify this space with the polish space of c\`{a}dl\`{a}g paths on $[0,1]$
with values in $S$.

Let $(N_{t})_{t\geq0}$ be a standard Poisson process with $N_{0}=0$ and jump
intensity $d-1$ defined on and adapted to $(\Omega,\mathcal{F},(\mathcal{F}%
_{t})_{0\leq t\leq1},P)$. We denote by $(T_{m})_{m\geq0}$ the jumping times of
$(N_{t})_{0\leq t\leq1}$, i.e. $T_{m}:=\inf\{0\leq t\leq1,$ $N_{t}\geq m\}$,
where the infimum over the empty set equals infinity. We shall denote by
$(X_{t})_{t\geq0}$ the c\`{a}dl\`{a}g-process starting in a uniformly
distributed way at points in $S$, i.e. $P(X_{0}=i)=\frac{1}{d}$, and having
the properties,%
\begin{align}
T_{m}  &  =\inf\{t\geq T_{m-1}\text{ with }X_{t}\neq X_{T_{m-1}}%
\}\label{process1}\\
P(X_{T_{m}}  &  =k|X_{T_{m}-}=l)=\frac{1}{d-1} \label{process2}%
\end{align}
for $m\geq1$ and $k\neq l\in\{1,\dots,d\}$, i.e. the process jumps at
Poissonian jumping times $T_{m}$ in a uniformly distributed way to another
state. Note that this process is stationary, i.e. $P(X_{t}=i)=\frac{1}{d}$ for
each $0\leq t\leq1$ and $1\leq i\leq d$.

We define for $0\leq t\leq1$%
\[
Z_{t}:=a_{X_{0}X_{T_{1}}}a_{X_{T_{1}}X_{T_{2}}}\dots a_{X_{T_{n-1}}%
X_{T_{N_{t}}}}=\prod_{i=1}^{N_{t}}a_{X_{T_{i-1}}X_{T_{i}}}.
\]
where the product is almost surely well defined as $N_{1}<\infty$ almost
surely. The empty product is defined to be $1$. We set%
\[
Y_{t}:=(Z_{t},X_{t})
\]
for $0\leq t\leq1$. Then $(Y_{t})_{0\leq t\leq1}:=(Z_{t},X_{t})_{0\leq t\leq
1}$ is a process with c\`{a}dl\`{a}g paths starting in a uniformly distributed
way at $\{(1,1),\dots,(1,d)\}$ in $S$.

\begin{remark}
We may and will choose the polish space of c\`{a}dl\`{a}g paths on $[0,1]$
with values in $S$ as probability space $\Omega$ with the Borel probability
measure $P$, such that the coordinate process on $\Omega$ together with the
canonical filtrations $(\mathcal{F}_{t})_{0\leq t\leq1}$ satisfy the above
requirements. Hence the process is a well defined map on the \emph{entire}
probability space $\Omega$, which will allow us to leave out the usual
''almost surely'' at several occassions.
\end{remark}

We define probability measures $P^{i}$ on $(\Omega,\mathcal{F},(\mathcal{F}%
_{t})_{0\leq t\leq1})$ by conditioning on the event $X_{0}=i$ for
$i\in\{1,\dots,d\}$, i.e. $P^{i}:=P(.|X_{0}=i)$. With respect to the
probability measures $P^{i}$ we define, for $(\zeta,i)\in S$, a process
$(Y_{t}^{(\zeta,i)})_{0\leq t\leq1}:=(Z_{t}^{(\zeta,i)},X_{t}^{i})_{0\leq
t\leq1}$ on $(\Omega,\mathcal{F},(\mathcal{F}_{t})_{0\leq t\leq1},P^{i})$
through%
\begin{align*}
X_{t}^{i}  &  =X_{t}\\
Z_{t}^{(\zeta,i)}  &  =\zeta Z_{t}%
\end{align*}
for $0\leq t\leq1$.

\begin{proposition}
Let $A$ be a hermitian $d\times d$ matrix. The family of processes
$(Y_{t}^{(\zeta,i)})_{0\leq t\leq1}$ on $(\Omega,\mathcal{F},(\mathcal{F}%
_{t})_{0\leq t\leq1},P^{i})$ for $(\zeta,i)\in S$ defines a Markov process
with generator%
\[
\mathcal{A}f(\zeta,i)=\sum_{\substack{j=1\\j\neq i}}^{d}(f(\zeta
a_{ij},j)-f(\zeta,i)).
\]
for all $f\in C(S,\mathbb{C})$ and $(\zeta,i)\in S$, where $C(S,\mathbb{C})$
denotes the set of continuous functions on $S$.
\end{proposition}

\begin{proof}
Fix $f\in C(S,\mathbb{C})$ and $(\zeta,i)\in S$, then%
\begin{align*}
&  \frac{1}{t}E_{P^{i}}(f(Z_{t}^{(\zeta,i)},X_{t}^{i})-f(\zeta,i))\\
&  =\frac{1}{t}\sum_{k=0}^{\infty}E_{P^{i}}(f(Z_{t}^{(\zeta,i)},X_{t}%
^{i})-f(\zeta,i)|N_{t}=k)P(N_{t}=k)\\
&  =\frac{1}{t}\frac{1}{d-1}\sum_{\substack{j=1\\j\neq i}}^{d}(f(\zeta
a_{ij},j)-f(\zeta,i))\frac{(d-1)t}{1!}e^{-(d-1)t}+\frac{1}{t}O(t^{2})\\
&  \rightarrow\sum_{\substack{j=1\\j\neq i}}^{d}(f(\zeta a_{ij},j)-f(\zeta
,i)),
\end{align*}
as $t\rightarrow0$.
\end{proof}

The Feynman-Kac formula allows for a stochastic interpretation of the
BMV-measure $\mu^{A,B}$. Fix $r\in\mathbb{C}^{d}$, then Theorem \ref{fk2}
asserts for functions $f^{r}(\zeta,i):=r_{i}\zeta$, for $(\zeta,i)\in S$, the
following formula to calculate $\exp(t(A-zB))r$.

\begin{corollary}
Given $r\in\mathbb{C}^{d}$, a hermitian $d\times d$ matrix $A$ and a diagonal
matrix $B$ with non-negative entries $b_{1},\dots,b_{n}$, we have%
\begin{gather*}
E\left(  \exp(\int_{0}^{1}a(X_{s})ds-z\int_{0}^{1}b(X_{s})ds)f^{r}%
(Y_{1})|X_{0}=i\right) \\
=E_{P^{i}}\left(  \exp(\int_{0}^{1}a(X_{s}^{i})ds-z\int_{0}^{1}b(X_{s}%
^{i})ds)f^{r}(Y_{1}^{(1,i)})\right) \\
=e^{-(d-1)}(\exp(A-zB)r)_{i}%
\end{gather*}
holds true for all $z\in\mathbb{C}$ and $1\leq i\leq d$. Here $a=(a_{1}%
,\dots,a_{n})$ denotes the (real) vector of diagonal elements of
$A$.\label{fk3}
\end{corollary}

This formula allows for an interpretation of $z\mapsto(\exp(A-zB)r)_{i}$ as
Laplace transform of the random variable $\int_{0}^{1}b(X_{s})ds$ under the
(signed) measure $Q^{i}$ on $\Omega$%
\[
\frac{dQ^{i}}{dP}=\frac{1}{P(X_{0}=i)}\exp(\int_{0}^{1}a(X_{s})ds)f^{r}%
(Y_{1})1_{\{X_{0}=i\}},
\]
since $\int_{0}^{1}b(X_{s})ds$ appears linearly in $-z$ in the exponent.

Given a hermitian $d\times d$ matrix $A$ and a diagonal matrix $B$ with
non-negative diagonal entries $b_{1},\dots,b_{d}$, we define an $\mathcal{F}%
_{0}$-measurable random variable $f$ on $\mathbb{C}\times\{1,\dots
,d\}\times(\Omega,\mathcal{F}_{0},P)$ to obtain a closed formula for the trace
$z\mapsto\operatorname*{tr}(\exp(A-zB))$, namely%
\begin{equation}
f(\zeta,i):=de^{d-1}\zeta\left\{
\begin{array}
[c]{c}%
1\text{ if }X_{0}=i\\
0\text{ if }X_{0}\neq i
\end{array}
\right.  \label{def-f}%
\end{equation}
for $\zeta\in\mathbb{C}$ and $i=1,\dots,d$. By Corollary \ref{fk3} and
Definition (\ref{def-f}) we obtain, using the notation of Theorem \ref{TH},%
\begin{align}
\phi^{A,B}(z)  &  =\sum_{i=1}^{d}\left\langle e_{i},\exp(A-zB)e_{i}%
\right\rangle \nonumber\\
&  =\sum_{i=1}^{d}E_{P^{i}}(\exp(\int_{0}^{1}a(X_{s}^{i})ds-z\int_{0}%
^{1}b(X_{s}^{i})ds)f^{e_{i}}(Y_{1}))e^{d-1}d\frac{1}{d}\nonumber\\
&  =\sum_{i=1}^{d}E(\exp(\int_{0}^{1}a(X_{s})ds-z\int_{0}^{1}b(X_{s}%
)ds)f(Y_{1})|X_{0}=i)P(X_{0}=i)\nonumber\\
&  =E\left(  \exp(\int_{0}^{1}a(X_{s})ds-z\int_{0}^{1}b(X_{s})ds)f(Y_{1}%
)\right)  \label{fk4}%
\end{align}
for $z\in\mathbb{C}$.

We now derive a series representation of the measure $\mu^{A,B}$. The function
$f(Y_{1})$ can take non-zero values only at loops, i.e. $X_{0}=X_{1}$. First
we introduce the subset $\Omega_{n}\subset\Omega$ consisting of those paths
which form \textbf{loops} in $\{1,\dots,d\}$ on $[0,1]$ with precisely $n$
jumps for $n\geq2$, i.e. $\Omega_{n}:=\{X_{0}=X_{1},N_{1}=n\}$. We define the
set $C_{n}\subset\{1,\dots,d\}^{n}$ as image set of the path random variable%
\begin{align*}
p_{n}  &  :\Omega_{n}\rightarrow\{1,\dots,d\}^{n}\\
\omega &  \mapsto(X_{T_{1}-}(\omega),\dots,X_{T_{n}-}(\omega))
\end{align*}
So the subset $C_{n}$ of $\{1,\dots,d\}^{n}$ is characterized as set of all
$n$-tuples such that no neighbors are equal and the last element $X_{T_{n}-}$
is different from the first one $X_{T_{1}-}=X_{0}$. We denote $C:=\cup
_{n\geq0}C_{n}$. Elements $\gamma\in C_{n}$ are called \textbf{favorable
paths} of length $n$.

The map $\operatorname{ord}$ associates to $\gamma\in C_{n}$ a monomial in the
variables $a_{ij}$, which is called \textbf{order of the path}. The quantities
$l_{ij}(\gamma)$ are the respective powers of $a_{ij}$ in the monomial
$\operatorname{ord}(\gamma)$: for $\gamma\in C_{n}$ we define%
\begin{align}
\operatorname{ord}(\gamma)  &  :=a_{\gamma_{1}\gamma_{2}}a_{\gamma_{2}%
\gamma_{3}}\dots a_{\gamma_{n-1}\gamma_{n}}a_{\gamma_{n}\gamma_{1}%
}\label{order}\\
&  =\prod_{i<j}a_{ij}^{l_{ij}(\gamma)}.
\end{align}
The \textbf{characteristic }$\operatorname{char}(\gamma)=(k_{1}(\gamma
),\dots,k_{d}(\gamma))$ of a path $\gamma\in C_{n}$ is defined by the number
$k_{j}(\gamma)$ of visits in state $j$%
\[
k_{j}(\gamma):=\#\{l\text{ such that }\gamma_{l}=j\}.
\]
Notice that the following formula holds for $\gamma\in C_{n}$,%
\begin{equation}
\frac{1}{2}\sum_{j\neq i}l_{ij}(\gamma)=k_{i}(\gamma), \label{l-k-relation}%
\end{equation}
which leads in dimension $2$ and $3$ to one-to-one relations between
$\operatorname{char}(\gamma)$ and $\operatorname{ord}(\gamma)$ (see Lemma
\ref{charjumps}).

We shall denote by $\Delta_{n}$ the $n$-simplex in $\mathbb{R}^{n+1}$, i.e.
the set of vectors $(t_{1},\dots,t_{n+1})\in\mathbb{R}^{n+1}$ with $\sum
_{i=1}^{n+1}t_{i}=1$ and $t_{i}\geq0$. On the $n$-simplex we shall consider
the normalized uniform law $\lambda_{n}$, i.e. $\lambda_{n}(\Delta_{n})=1$.
For $\gamma\in C_{n}$, we consider the set $p_{n}^{-1}(\gamma)\subset
\Omega_{n}$. On $p_{n}^{-1}(\gamma)$ we consider the conditional probability
$P_{\gamma}:=P(.|p_{n}=\gamma)$ and the random variable%
\[
\operatorname*{dur}:p_{n}^{-1}(\gamma)\rightarrow\Delta_{n}%
\]
via $\operatorname*{dur}:=(T_{1},T_{2}-T_{1},\dots,T_{n}-T_{n-1},1-T_{n})$,
which has a uniform distribution $\lambda_{n}$ on the simplex $\Delta_{n}$
under $P_{\gamma}$. Indeed since the conditional distribution of $(T_{1}%
,T_{2}-T_{1},T_{3}-T_{2},\dots,T_{n}-T_{n-1})$ for $n\geq1$ under the
condition $N_{1}=n$ is uniform, we can conclude the result, i.e.%
\begin{gather*}
P_{\gamma}(T_{1}\in\lbrack t_{1},t_{1}+dt_{1}],T_{2}-T_{1}\in\lbrack
t_{2},t_{2}+dt_{2}],\dots,T_{n}-T_{n-1}\in\lbrack t_{n},t_{n}+dt_{n}])\\
=\frac{1}{\frac{1}{d}\frac{e^{-(d-1)}}{n!}}\frac{1}{d}\frac{1}{d-1}%
dt_{1}(d-1)e^{-t_{1}(d-1)}\dots\frac{1}{d-1}dt_{n}(d-1)e^{-t_{n}%
(d-1)}e^{-(1-\sum_{i=1}^{n}t_{i})(d-1)}\\
=\frac{dt_{1}\dots dt_{n}e^{-(d-1)}(d-1)^{n}\frac{1}{d}\frac{1}{(d-1)^{n}}%
}{\frac{1}{d}\frac{e^{-(d-1)}}{n!}}=n!dt_{1}\dots dt_{n}%
\end{gather*}
for $\sum_{i=1}^{n}t_{i}\leq1$. Here we apply that set $p_{n}^{-1}(\gamma)$
for $\gamma\in C_{n}$ has probability
\begin{equation}
P(p_{n}^{-1}(\gamma))=\frac{1}{d(d-1)^{n}}\frac{e^{-(d-1)}(d-1)^{n}}{n!}%
=\frac{1}{d}\frac{e^{-(d-1)}}{n!}, \label{prob-n-jumps}%
\end{equation}
since the probability for a trajectory to have $n$ jumps is $\frac
{e^{-(d-1)}(d-1)^{n}}{n!}$.

On the simplex $\Delta_{n}$ we define for a vector $h\in\mathbb{R}^{n+1}$ the
real-valued random variable $\mathcal{Y}^{h}=\sum_{i=1}^{n+1}t_{i}h_{i}$, and
for $g\in\mathbb{R}^{n+1}$ the measure $Q^{g}$ with%
\[
\frac{dQ^{g}}{d\lambda_{n}}:=\exp(\sum_{i=1}^{n+1}t_{i}g_{i})
\]
with respect to the uniform distribution on $\Delta_{n}$. The image measure of
$Q^{g}$ under $\mathcal{Y}^{h}$ is denoted by $\eta^{g;h}$, which is a measure
on $\mathbb{R}$ with support in the convex hull of $h_{1},\dots,h_{n+1}$.

\begin{theorem}
Let $A$ be a hermitian matrix, $B$ a diagonal matrix with non-negative,
mutually different entries $b_{1},\dots,b_{d}$. The diagonal elements of $A$
are denoted by $a_{1},\dots,a_{d}$. Then the measure (see Definition
\ref{bmvdef}) $\mu^{A,B}$ is a signed measure decomposing into an absolutely
continuous and singular part%
\begin{equation}
\mu^{A,B}(dx)=\sum_{i=1}^{d}\exp(a_{i})\delta_{b_{i}}(dx)+\psi^{A,B}(x)dx.
\label{bmvmeasure}%
\end{equation}
$\psi^{A,B}$ is a piecewise continuous function with possible disconituities
at $b_{i}$ and with support in $[\min_{i}b_{i},\max b_{i}]$. We have%
\begin{equation}
\psi^{A,B}(x)=\sum_{\gamma\in C}\phi(\gamma,x)a_{\gamma_{1}\gamma_{2}}\dots
a_{\gamma_{n}\gamma_{1}}, \label{seriesformu}%
\end{equation}
where the density $\phi(\gamma,x)$ is defined by%
\begin{equation}
\phi(\gamma,x)dx:=\frac{1}{n!}\eta^{a_{\gamma_{1}},\dots,a_{\gamma_{n}%
},a_{\gamma_{1}};b_{\gamma_{1}},\dots,b_{\gamma_{n}},b_{\gamma_{1}}}(dx)
\label{densitydimd}%
\end{equation}
for $\gamma\in C_{n}$.\label{dyson}
\end{theorem}

\begin{proof}
We can decompose the Feynman-Kac formula (\ref{fk4}) by the number of jumps
$N_{1}$, which appear up to time $1$. This leads to%
\begin{align*}
&  E(\exp(\int_{0}^{1}a(X_{s})ds-z\int_{0}^{1}b(X_{s})ds)f(Y_{1}))\\
&  =\sum_{n=0}^{\infty}E(\exp(\int_{0}^{1}a(X_{s})ds-z\int_{0}^{1}%
b(X_{s})ds)f(Y_{1}),N_{1}=n),
\end{align*}
wherefrom we obtain the basic decomposition of regular and singular part. The
probability of taking $0$ jumps up to time $1$ and starting at $i$ is
$\frac{e^{-(d-1)}}{d}$, which leads to the expression for the singular part of
$\mu^{A,B}$ in (\ref{bmvmeasure}) by the definition of $f$. Again by the
definition of $f$ , formula (\ref{fk4}) and formula (\ref{prob-n-jumps}) we
obtain that, for $n\geq1$,%
\begin{gather*}
E(\exp(\int_{0}^{1}a(X_{s})ds-z\int_{0}^{1}b(X_{s})ds)f(Y_{1}),N_{1}=n)\\
=\sum_{\gamma\in C_{n}}E(\exp(\int_{0}^{1}a(X_{s})ds-z\int_{0}^{1}%
b(X_{s})ds)f(Y_{1})|p_{n}=\gamma)P(p_{n}=\gamma)\\
=\sum_{\gamma\in C_{n}}E_{P_{\gamma}}(\exp(\int_{0}^{1}a(X_{s}|_{p_{n}%
^{-1}(\gamma)})ds-z\int_{0}^{1}b(X_{s}|_{p_{n}^{-1}(\gamma)})ds)f(Y_{1}%
|_{p_{n}^{-1}(\gamma)}))P(p_{n}=\gamma)\\
=\sum_{\gamma\in C_{n}}\int_{0}^{\infty}\exp(-zx)\frac{1}{n!}\eta
^{a_{\gamma_{1}},\dots,a_{\gamma_{n}},a_{\gamma_{1}};b_{\gamma_{1}}%
,\dots,b_{\gamma_{n}},b_{\gamma_{1}}}(dx)\operatorname{ord}(\gamma)
\end{gather*}
holds true. We have applied that $f$ equals $de^{(d-1)}\operatorname{ord}%
(\gamma)$ on the set $p_{n}^{-1}(\gamma)$, since this set consists of loops
with favorable path $\gamma$ and $Z_{1}$ takes precisely the value
$\operatorname{ord}(\gamma)$. In addition $de^{(d-1)}P(p_{n}^{-1}%
(\gamma))=\frac{1}{n!}$.

We have to show that the measure $\eta^{a_{\gamma_{1}},\dots,a_{\gamma_{n}%
},a_{\gamma_{1}};b_{\gamma_{1}},\dots,b_{\gamma_{n}},b_{\gamma_{1}}}(dx)$ has
a density, which can be seen from the fact that the intersection of
$\Delta_{n}$ and the set $\{\sum_{i=1}^{n}b_{\gamma_{i}}t_{i}+b_{\gamma_{1}%
}t_{n+1}=x\}$ depends in a differentiable way on $x$. The sum starts with
$n=2$ since there are no loops with only one jump.
\end{proof}

\begin{example}
We illustrate the stochastic approach by the case $d=2$. Since a loop
$\gamma\in C_{n}$ only appears if $n$ is even and has the form $121\dots$ or
$212\dots$, the contributions in the above series are necessarily
non-negative: indeed for a hermitian $2\times2$ matrix $A$ we must have that
$\operatorname{ord}(\gamma)\geq0$ for all loops $\gamma$ and the measures
$\eta$ are non-negative either. Hence the density is non-negative. Again we
note that the validity of the BMV-conjecture for $d=2$ is well-known (see e.g.
\cite{BesMouVil:75}).
\end{example}

\begin{remark}
Theorem \ref{dyson} can also be derived from the well-known Dyson expansion
(see for instance \cite{FanWer:00}). Still we believe that the stochastic
reasoning underlying our proof has special merits: it leads us to a
probabilistic and combinatorial point of view (see Section 3 ''Stochastic Semantics'').
\end{remark}

\begin{remark}
We have formulated Theorem \ref{dyson} for Hermitian matrices $A$ as this is
presently our natural framework. But it is clear that it may as well be
formulated for general $d\times d$ matrices $A$.
\end{remark}

\section{Stochastic Semantics}

From now on we assume $d=3$ and $b_{1}=0$, we shall write $a_{i}=a_{ii}$ for
$i=1,2,3$. In particular all matrices will be real from now on. From the point
of view of stochastic processes we now have a dynamic picture of the problem
to calculate the measure $\mu^{A,B}$: we consider paths with values in the set
$\{1,2,3\}$, which are favorable in the sense that two neighboring elements
are different and the last element is different from the first one. The
combinatorics of these paths leads us to a particular way to sum up the series
(\ref{seriesformu}). We think about trajectories dynamically as loops on the
vertices $\{e_{1},e_{2},e_{3}\}$ of the $2$-simplex. The total times $\xi_{i}%
$, which the trajectory stays in state $i$ during time $[0,1]$, form an
element of the $2$-simplex. Only if $\sum_{i=1}^{3}b_{i}\xi_{i}\in\lbrack
x,x+dx]$, the trajectory contributes to the density $\psi^{A,B}(x)dx$.

We now fix $0<b_{3}\leq b_{2}$ and $x\in]0,b_{3}[$. Due to the following
choice of parameters we choose the unsual convention $b_{3}\leq b_{2}$. The
intersection of $\Delta_{2}$ with $\sum_{i=1}^{3}b_{i}\xi_{i}=x$ will be
parametrized by%
\[
t\mapsto((1-x_{2})+t(x_{2}-x_{3}),x_{2}(1-t),x_{3}t)
\]
with real numbers $0<x_{2}\leq x_{3}<1$ and $t\in\lbrack0,1]$. We shall denote
this line segment by $L^{x_{2},x_{3}}$ and we obtain the relations%
\begin{align*}
b_{2}x_{2}  &  =x,\\
b_{3}x_{3}  &  =x.
\end{align*}
In particular we observe that -- for given $x$ -- the numbers $b_{2},b_{3}$
and $x_{2},x_{3}$ determine each other. In order to obtain $x_{2}\leq x_{3}$
we have been choosing $b_{3}\leq b_{2}$.

We apply the notions of the previous section. The \textbf{characteristic}
$\operatorname{char}(\gamma)=(k_{1}(\gamma),k_{2}(\gamma),k_{3}(\gamma))$ of a
path $\gamma\in C$ is the number of visits in the points $1,2,3$. Clearly
$k_{1}+k_{2}+k_{3}=n$. We shall observe in the following Lemma that in
dimension $3$ the characteristic already determines the number of jumps
between $1-2,1-3$ and $2-3$, denoted by $l_{12},l_{13}$ and $l_{23}$. These
quantities are defined via%
\[
a_{12}^{l_{12}(\gamma)}a_{13}^{l_{13}(\gamma)}a_{23}^{l_{23}(\gamma
)}:=a_{\gamma_{1}\gamma_{2}}a_{\gamma_{2}\gamma_{3}}\dots a_{\gamma
_{n-1}\gamma_{n}}a_{\gamma_{n}\gamma_{1}}=\operatorname{ord}(\gamma)
\]
and the numbers $l_{ij}(\gamma)$ of jumps between $i$ and $j$ only depend on
$\operatorname{char}\emph{(}\gamma)=(k_{1}(\gamma),k_{2}(\gamma),k_{3}%
(\gamma))$ for $\gamma\in C_{n}$.

\begin{lemma}
The characteristic $\operatorname{char}(\gamma)=(k_{1}(\gamma),k_{2}%
(\gamma),k_{3}(\gamma))$ of a path $\gamma\in C$ and the powers $(l_{12}%
(\gamma),l_{13}(\gamma),l_{23}(\gamma))$ of the order $\operatorname{ord}%
(\gamma)$ are in one-to-one relation. By abuse of notation we may therefore
write $l_{ij}(\gamma)=l_{ij}(k_{1}(\gamma),k_{2}(\gamma),k_{3}(\gamma
))$.\label{charjumps}
\end{lemma}

\begin{proof}
We take formula (\ref{l-k-relation}) and solve it for $l_{ij}$ given
$\operatorname{char}(\gamma)$, we obtain%
\begin{align*}
l_{12}  &  =k_{1}+k_{2}-k_{3}\\
l_{13}  &  =k_{1}+k_{3}-k_{2}\\
l_{23}  &  =k_{2}+k_{3}-k_{1},
\end{align*}
which yields the result.
\end{proof}

Next we calculate in our particular setting (recall that $d=3$ and $b_{1}=0$)
explicitly the density of $\eta^{a_{\gamma_{1}},\dots,a_{\gamma_{n}}%
,a_{\gamma_{1}};b_{\gamma_{1}},\dots,b_{\gamma_{n}},b_{\gamma_{1}}}$ at $x$.

\begin{lemma}
For $k_{1},k_{2},k_{3}\geq0$ define a probability density $f$ on $\Delta_{2}$
(this time with respect to uniform distribution $\frac{1}{2}\lambda_{2}$ of
total mass $\frac{1}{2}$) given through%
\begin{equation}
f(\xi_{1},\xi_{2},\xi_{3})=\beta(k_{1},k_{2},k_{3})\xi_{1}^{k_{1}-1}\xi
_{2}^{k_{2}-1}\xi_{3}^{k_{3}-1}\exp(a_{1}\xi_{1}+a_{2}\xi_{2}+a_{3}\xi_{3}),
\label{simplex-density}%
\end{equation}
where%
\[
\beta(k_{1},k_{2},k_{3})=\frac{(k_{1}+k_{2}+k_{3}-1)!}{(k_{1}-1)!(k_{2}%
-1)!(k_{3}-1)!}%
\]
for $k_{i}\geq1$. We fix a path $\gamma\in C$ with characteristic
$\operatorname{char}(\gamma)=(k_{1},k_{2},k_{3})$, $n:=k_{1}+k_{2}+k_{3}$, and
define $\gamma^{\prime}=(\gamma_{1},\dots,\gamma_{n},\gamma_{1})\in
\{1,\dots,d\}^{n+1}$ and%
\[
pr_{\gamma}:\Delta_{n}\rightarrow\Delta_{2}%
\]
through $\xi_{i}(\gamma):=(pr_{\gamma}(t_{1},\dots,t_{n+1}))_{i}%
=\sum_{\substack{j=1\\\gamma_{j}^{\prime}=i}}^{n+1}t_{j}$ for $i=1,2,3$.
Notice that the law of the real-valued random variable $\omega\mapsto b_{2}%
\xi_{2}(p_{n}(\omega))+b_{3}\xi_{3}(p_{n}(\omega))$ under $P_{\gamma}$ is
$\eta^{a_{\gamma_{1}},\dots,a_{\gamma_{n}},a_{\gamma_{1}};b_{\gamma_{1}}%
,\dots,\gamma_{n},\gamma_{1}}(dx)=n!\phi(\gamma,x)dx$ (see formula
\ref{densitydimd}) Then the following assertions hold true:

\begin{enumerate}
\item Assume $k_{i}\geq1$, for $i=1,2,3$, then the law of the random variable
$pr_{\gamma}$ has density
\begin{equation}
(k_{1}+k_{2}+k_{3})\frac{\xi_{\gamma_{1}}}{k_{\gamma_{1}}}f(\xi_{1},\xi
_{2},\xi_{3})=n\frac{\xi_{\gamma_{1}}}{k_{\gamma_{1}}}f(\xi_{1},\xi_{2}%
,\xi_{3}) \label{formuladensitydimd}%
\end{equation}
with respect to the measure $\frac{1}{2}\lambda_{2}$ on $\Delta_{2}$. Notice
that the state appearing in $\gamma_{1}$ is counted twice in the density.

\item For $k_{i}\geq1$, $i=1,2,3$, and $x\in$ $]0,b_{3}[$%
\begin{align}
\phi(\gamma,x)  &  =\frac{1}{(n-1)!}\sqrt{\frac{1}{b_{2}b_{3}}}\int_{0}%
^{1}\left\{
\begin{array}
[c]{c}%
\frac{(1-x_{2})+t(x_{2}-x_{3})}{k_{1}}\\
\frac{x_{2}(1-t)}{k_{2}}\\
\frac{x_{3}t}{k_{3}}%
\end{array}
\right\} \label{phigeneral}\\
&  f(1-x_{2})+t(x_{2}-x_{3}),x_{2}(1-t),x_{3}t)\sqrt{x_{2}x_{3}}dt
\end{align}
at $0<x<b_{3}$, where the cases in $\{\}$ pertain to $\gamma_{1}=1,2,3$.
Notice the relations $b_{2}x_{2}=b_{3}x_{3}=x$.

\item Assume $k_{1}=0$, and $x\in]0,b_{3}[$, then $\phi(\gamma,x)=0$ for all
$n\geq2$.

\item Assume $k_{2}=0$, and $x\in]0,b_{3}[$, then
\begin{equation}
\phi(\gamma,x)=\frac{1}{(n-1)!}\frac{(k_{1}+k_{3}-1)!}{(k_{1}-1)!(k_{3}%
-1)!}\left\{
\begin{array}
[c]{c}%
\exp(a_{1}(1-x_{3})+a_{3}x_{3})\frac{(1-x_{3})^{k_{1}}x_{3}^{k_{3}-1}}%
{k_{1}b_{3}}\text{ if }\gamma_{1}=1\\
\exp(a_{1}(1-x_{3})+a_{3}x_{3})\frac{(1-x_{3})^{k_{1}-1}x_{3}^{k_{3}}}%
{k_{3}b_{3}}\text{ if }\gamma_{1}=3
\end{array}
\right.  \label{phi2null}%
\end{equation}
and $n$ is necessarily even.

\item Assume $k_{3}=0$, and $x\in]0,b_{3}[$, then
\begin{equation}
\phi(\gamma,x)=\frac{1}{(n-1)!}\frac{(k_{1}+k_{2}-1)!}{(k_{1}-1)!(k_{2}%
-1)!}\left\{
\begin{array}
[c]{c}%
\exp(a_{1}(1-x_{2})+a_{2}x_{2})\frac{(1-x_{2})^{k_{1}}x_{2}^{k_{2}-1}}%
{k_{1}b_{2}}\text{ if }\gamma_{1}=1\\
\exp(a_{1}(1-x_{2})+a_{2}x_{2})\frac{(1-x_{2})^{k_{2}-1}x_{2}^{k_{2}}}%
{k_{2}b_{2}}\text{ if }\gamma_{1}=2
\end{array}
\right.  \label{phi3null}%
\end{equation}
and $n$ is necessarily even.\label{originaldensity}
\end{enumerate}
\end{lemma}

\begin{proof}
Fix $\gamma\in C_{n}$ and let $\gamma^{\prime}=(\gamma_{1},\dots,\gamma
_{n},\gamma_{1})$. We first set $a_{i}=0$ for $i=1,2,3$. By direct computation
we verify that now $f$ indeed defines a probability measure on $\Delta_{2}$,
hence the norming factor is correct (the actual form of $f$ stems from pushing
forward with $pr_{\gamma}$ and simply observing that a sum of independent
uniformly distributed variables leads to a $\beta$-distribution). In the chart
$\pi_{12}$ (projection from $\Delta_{2}$ on the first two components in
$\mathbb{R}^{3}$) the volume element $\frac{1}{2}\lambda_{2}(d\xi)$ equals
$d\xi_{1}d\xi_{2}$ on $\{(\xi_{1},\xi_{2})\in\mathbb{R}^{2},\quad\xi_{1}%
,\xi_{2}\geq0,\ \xi_{1}+\xi_{2}\leq1\}$.%
\begin{gather*}
\frac{1}{2}\int_{\Delta_{2}}f(\xi_{1},\xi_{2},\xi_{3})\lambda_{2}(d\xi
)=\beta(k_{1},k_{2},k_{3})\int_{0}^{1}\int_{0}^{1-\xi_{2}}\xi_{1}^{k_{1}-1}%
\xi_{2}^{k_{2}-1}(1-\xi_{1}-\xi_{2})^{k_{3}-1}d\xi_{1}d\xi_{2}\\
=\beta(k_{1},k_{2},k_{3})\int_{0}^{1}\xi_{2}^{k_{2}-1}(1-\xi_{2})^{k_{1}%
+k_{3}-1}\int_{0}^{1-\xi_{2}}\frac{\xi_{1}^{k_{1}-1}}{(1-\xi_{2})^{k_{1}-1}%
}(1-\frac{\xi_{1}}{1-\xi_{2}})^{k_{3}-1}d(\frac{\xi_{1}}{1-\xi_{2}})d\xi_{2}\\
=\beta(k_{1},k_{2},k_{3})\int_{0}^{1}\xi_{2}^{k_{2}-1}(1-\xi_{2})^{k_{1}%
+k_{3}-1}\int_{0}^{1}\eta^{k_{1}-1}(1-\eta)^{k_{3}-1}d\eta d\xi_{2}\\
=\beta(k_{1},k_{2},k_{3})\frac{(k_{2}-1)!(k_{1}+k_{3}-1)!}{(k_{1}+k_{2}%
+k_{3}-1)!}\frac{(k_{1}-1)!(k_{3}-1)!}{(k_{1}+k_{3}-1)!}=1.
\end{gather*}
We continue now with general $a_{i}$. Calculating the formula of the density
$\phi(\gamma,x)$ at $x\in]0,b_{3}[$ amounts to calculating the mass of
$pr_{\gamma}$ passed by the line $L^{x_{2},x_{3}}$ through variations in $x$.
Fixing $b_{2},b_{3}$ we thus fix $x_{2},x_{3}$. The area of the quadrangle
with corners at $e_{1}+x_{i}(e_{i}-e_{1})$, $e_{1}+(x_{i}+dx_{i})(e_{i}%
-e_{1})$, for $i=2,3$, with respect to the measure $\frac{1}{2}\lambda
_{2}(d\xi)$ -- under a small variation $dx$ of $x$ -- is given by%
\begin{align*}
\frac{1}{2}(x_{3}dx_{2}+x_{2}dx_{3})  &  =\frac{1}{b_{3}b_{2}}xdx\\
&  =\frac{1}{b_{3}b_{2}}\sqrt{b_{2}b_{3}x_{2}x_{3}}dx\\
&  =\sqrt{\frac{1}{b_{2}b_{3}}}\sqrt{x_{2}x_{3}}dx.
\end{align*}
Shrinking the side $\operatorname*{conv}\{e_{1}+x_{2}(e_{2}-e_{1}),e_{1}%
+x_{3}(e_{3}-e_{1})\}$ to an infitesimal element at the point $((1-x_{2}%
)+t(x_{2}-x_{3}),x_{2}(1-t),x_{3}t)$, for $t\in\lbrack0,1]$, on $L^{x_{2}%
,x_{3}}$ leads to the appropriate area element%
\[
\sqrt{\frac{1}{b_{2}b_{3}}}\sqrt{x_{2}x_{3}}dxdt.
\]
Hence we can determine $\phi_{n}(\gamma,x)$ through equation
(\ref{densitydimd}) and formula (\ref{formuladensitydimd}) evaluated at
$((1-x_{2})+t(x_{2}-x_{3}),x_{2}(1-t),x_{3}t)$, for $t\in\lbrack0,1]$,%
\begin{gather*}
P_{\gamma}(b_{2}\xi_{2}\circ p_{n}+b_{3}\xi_{3}\circ p_{n}\in\lbrack
x,x+dx])\\
=\sqrt{\frac{1}{b_{2}b_{3}}}\sqrt{x_{2}x_{3}}dx\frac{1}{(n-1)!}\beta
(k_{1},k_{2},k_{3})\int_{0}^{1}\left\{
\begin{array}
[c]{c}%
\frac{(1-x_{2})+t(x_{2}-x_{3})}{k_{1}}\\
\frac{x_{2}(1-t)}{k_{2}}\\
\frac{x_{3}t}{k_{3}}%
\end{array}
\right\} \\
f(1-x_{2}+t(x_{2}-x_{3}),x_{2}(1-t),x_{3}t)dt.
\end{gather*}

For the degenerate cases we perform the same program. We first calculate the
density of the law of $pr_{\gamma}$ if one of the $k_{i}$ is zero, which is a
density supported by one edge of the simplex $\Delta_{2}$. Assume $k_{3}=0$.
With respect to the uniform distribution with total mass $1$ on the edge
$\operatorname*{conv}\{e_{1},e_{2}\}$ of $\Delta_{2}$ we obtain for
$k_{1},k_{2}\geq1$%
\[
(k_{1}+k_{2})\frac{\xi_{\gamma_{1}}}{k_{\gamma_{1}}}\frac{(k_{1}+k_{2}%
-1)!}{(k_{1}-1)!(k_{2}-1)!}\xi_{1}^{k_{1}-1}\xi_{2}^{k_{2}-1}\exp(a_{1}\xi
_{1}+a_{2}\xi_{2})
\]
and similar for the other case. A small variation $dx$ in $x$ leads via
$\frac{dx}{b_{i}}=dx_{i}$ for $i=2,3$ to the desired results.
\end{proof}

In order to write the above densities in a more compact way we shall apply the
well-known formula%
\[
\frac{1}{\Gamma(\alpha)}\int_{0}^{1}g(t)t^{\alpha-1}\rightarrow g(0)
\]
as $\alpha\downarrow0$ for any continuous function $g:[0,1]\rightarrow
\mathbb{R}$. Hence we can apply $(k-1)!=\Gamma(k)$ for $k\geq0$ and obtain the
following proposition:

\begin{lemma}
For $\gamma\in C$ and $x\in]0,b_{3}[$, we obtain in the sense of
Gamma-functions%
\begin{align*}
\phi(\gamma,x)  &  =\frac{1}{(n-1)!}\sqrt{\frac{1}{b_{2}b_{3}}}\int_{0}%
^{1}\left\{
\begin{array}
[c]{c}%
\frac{(1-x_{2})+t(x_{2}-x_{3})}{k_{1}}\\
\frac{x_{2}(1-t)}{k_{2}}\\
\frac{x_{3}t}{k_{3}}%
\end{array}
\right\} \\
&  f(1-x_{2}+t(x_{2}-x_{3}),x_{2}(1-t),x_{3}t)\sqrt{x_{2}x_{3}}dt,
\end{align*}
for $\operatorname{char}(\gamma)=(k_{1},k_{2},k_{3})$, $k_{1}+k_{2}+k_{3}=n$,
and $k_{i}\geq0$, where the cases in $\{\}$ pertain to $\gamma_{1}=1,2,3$.
\end{lemma}

\begin{proof}
For $k_{i}\geq1$ there is nothing to prove. Assume now that we take the limit
$k_{2}\downarrow0$, hence $\gamma_{2}=1$ or $3$, since the vertex $2$ cannot
be starting point. We introduce furthermore
\begin{align*}
\lambda &  :=a_{1}(x_{2}-x_{3})-a_{2}x_{2}+a_{3}x_{3}\\
\mu &  :=a_{1}(1-x_{2})+a_{2}x_{2}%
\end{align*}
as in Remark \ref{remdiagonal2}. Hence the limit yields%
\begin{gather*}
\lim_{\alpha\downarrow0}\frac{1}{(n-1)!}\frac{(k_{1}+k_{3}-1)!}{(k_{1}%
-1)!(k_{3}-1)!}\sqrt{\frac{1}{b_{2}b_{3}}}\frac{1}{\Gamma(\alpha)}\int_{0}%
^{1}\left\{
\begin{array}
[c]{c}%
\frac{(1-x_{2})+t(x_{2}-x_{3})}{k_{1}}\text{ if }\gamma_{1}=1\\
\frac{x_{3}t}{k_{3}}\text{ if }\gamma_{1}=3
\end{array}
\right\} \\
((1-x_{2})+t(x_{2}-x_{3}))^{k_{1}-1}(x_{2}(1-t))^{\alpha-1}(x_{3}t)^{k_{3}%
-1}\sqrt{x_{2}x_{3}}dt\\
=\frac{\exp(\lambda+\mu)}{(n-1)!}\frac{(k_{1}+k_{3}-1)!}{(k_{1}-1)!(k_{3}%
-1)!}\frac{1}{x_{2}}\sqrt{\frac{x_{2}x_{3}}{b_{2}b_{3}}}\left\{
\begin{array}
[c]{c}%
\frac{(1-x_{3})^{k_{1}}x_{3}^{k_{3}-1}}{k_{1}}\text{ if }\gamma_{1}=1\\
\frac{(1-x_{3})^{k_{1}-1}x_{3}^{k_{3}}}{k_{3}}\text{ if }\gamma_{1}=3
\end{array}
\right. \\
=\frac{\exp(\lambda+\mu)}{(n-1)!}\frac{(k_{1}+k_{3}-1)!}{(k_{1}-1)!(k_{3}%
-1)!}\left\{
\begin{array}
[c]{c}%
\frac{(1-x_{3})^{k_{1}}x_{3}^{k_{3}-1}}{k_{1}b_{3}}\text{ if }\gamma_{1}=1\\
\frac{(1-x_{3})^{k_{1}-1}x_{3}^{k_{3}}}{k_{3}b_{3}}\text{ if }\gamma_{1}=3
\end{array}
\right.  ,
\end{gather*}
since $x_{2}b_{2}=x_{3}b_{3}=x$. Similarly for the third case.
\end{proof}

For the calculation of the BMV-measure $\mu^{A,B}$ we can make an essential
further simplification: it turns out that if we average over all paths
$\gamma$ with fixed characteristic $\operatorname{char}(\gamma)$ (and varying
the first entry $\gamma_{1}$) formulas (\ref{phigeneral})-(\ref{phi3null})
appear in a simpler form, which only depends on the characteristic. We define
the density%
\[
\chi(k_{1},k_{2},k_{3},x):=\frac{1}{\#\{\gamma\in C:\quad\operatorname{char}%
(\gamma)=(k_{1},k_{2},k_{3})\}}\sum_{\substack{\gamma\in
C\\\operatorname{char}(\gamma)=(k_{1},k_{2},k_{3})}}\phi(\gamma,x),
\]
i.e. the average of the densities $\phi(\gamma,x)$ where $\gamma$ ranges
through the paths with fixed characteristic $(k_{1},k_{2},k_{3})$.

\begin{lemma}
\label{Le3} We fix a path $\gamma\in C$ with characteristic
$\operatorname{char}(\gamma)=(k_{1},k_{2},k_{3})$ for $n\geq2$. Then the
following assertion holds,%
\begin{equation}
\chi(k_{1},k_{2},k_{3},x)=\sqrt{\frac{1}{b_{2}b_{3}}}\frac{1}{n!}\int_{0}%
^{1}f(1-x_{2}+t(x_{2}-x_{3}),x_{2}(1-t),x_{3}t)\sqrt{x_{2}x_{3}}dt
\label{chigeneral}%
\end{equation}

in the sense of Gamma-functions.\label{density}
\end{lemma}

\begin{remark}
\label{remdiagonal2}The exponential term in (\ref{simplex-density}) simplifies
to $e^{\lambda t+\mu}$ with $\lambda=a_{1}(x_{2}-x_{3})-a_{2}x_{2}+a_{3}x_{3}$
and $\mu=a_{1}(1-x_{2})+a_{2}x_{2}$. Hence we obtain, for a path with
characteristic $\operatorname{char}(\gamma)=(k_{1},k_{2},k_{3})$, $k_{i}%
\geq1,$ by the binomial theorem and the Beta integral that
\begin{align*}
&  \chi(k_{1},k_{2},k_{3},x)=\frac{(1-x_{2})^{k_{1}-1}x_{2}^{k_{2}}%
x_{3}^{k_{3}}}{nx}\\
&  \quad\times\sum_{L\geq0}\sum_{r=0}^{k_{1}-1}e^{\mu}\frac{\lambda^{L}}%
{L!}\binom{k_{1}-1}{r}\left(  \frac{x_{2}-x_{3}}{1-x_{2}}\right)  ^{r}%
\frac{(k_{3})_{L+r}}{(k_{1}-1)!(k_{2}+k_{3}+L+r-1)!},
\end{align*}
holds true, or --- by using $t=1-s$ --- an alternate representation,%
\begin{align*}
&  \chi(k_{1},k_{2},k_{3},x)=\frac{(1-x_{3})^{k_{1}-1}x_{2}^{k_{2}}%
x_{3}^{k_{3}}}{nx}\\
&  \quad\times\sum_{L\geq0}\sum_{r=0}^{k_{1}-1}\binom{k_{1}-1}{r}%
e^{\mu+\lambda}\frac{(-\lambda)^{L}}{L!}\left(  \frac{x_{3}-x_{2}}{1-x_{3}%
}\right)  ^{r}\frac{(k_{2})_{L+r}}{(k_{1}-1)!(k_{2}+k_{3}+L+r-1)!}.
\end{align*}
Here we apply the notion $(k)_{r}:=\frac{\Gamma(r+k)}{\Gamma(k)}$.
\end{remark}

\begin{proof}
[Proof of Lemma 3]For the proof we apply the representations of the densities
(\ref{phigeneral})-(\ref{phi3null}), and the fact that among all paths
$\gamma\in C_{n}$ with characteristic $\operatorname{char}(\gamma
)=(k_{1},k_{2},k_{3})$ the path with $\gamma_{1}=i$ appear with relative
frequency $\frac{k_{i}}{n}$, hence absolutely%
\[
\#\{\gamma\in C:\quad\operatorname{char}(\gamma)=(k_{1},k_{2},k_{3}%
)\}\frac{k_{i}}{n}%
\]
times. We calculate the density $\chi(k_{1},k_{2},k_{3},x)$ at $x\in]0,b_{3}%
[$. This leads for $k_{i}\geq1$ to%
\begin{align*}
\chi(k_{1},k_{2},k_{3},x)  &  =\frac{1}{%
\begin{array}
[c]{c}%
\#\{\gamma\in C:\quad\operatorname{char}\\
(\gamma)=(k_{1},k_{2},k_{3})\}
\end{array}
}\sum_{\substack{\gamma\in C\\\operatorname{char}(\gamma)=(k_{1},k_{2},k_{3}%
)}}\phi(\gamma,x)\\
&  =\frac{1}{(n-1)!}\sqrt{\frac{1}{b_{2}b_{3}}}\int_{0}^{1}\Bigl(\frac{k_{1}%
}{n}\frac{(1-x_{2})+t(x_{2}-x_{3})}{k_{1}}+\\
&  +\frac{k_{2}}{n}\frac{x_{2}(1-t)}{k_{2}}+\frac{k_{3}}{n}\frac{x_{3}t}%
{k_{3}}\Bigr)\\
&  f(1-x_{2}+t(x_{2}-x_{3}),x_{2}(1-t),x_{3}t)\sqrt{x_{2}x_{3}}dt\\
&  =\frac{1}{n!}\sqrt{\frac{1}{b_{2}b_{3}}}\int_{0}^{1}f(1-x_{2}+t(x_{2}%
-x_{3}),x_{2}(1-t),x_{3}t)\sqrt{x_{2}x_{3}}dt.
\end{align*}
For $k_{1}=0$ we conclude directly. For $k_{2}=0$ we use%
\begin{align*}
\chi(k_{1},0,k_{3},x)  &  =\frac{1}{(n-1)!}\frac{k_{1}}{n}\frac{(k_{1}%
+k_{3}-1)!}{k_{1}!(k_{3}-1)!}\frac{x_{3}^{k_{3}-1}(1-x_{3})^{k_{1}}}{b_{3}%
}\exp(a_{1}(1-x_{3})+a_{3}x_{3})+\\
&  +\frac{1}{(n-1)!}\frac{k_{3}}{n}\frac{(k_{1}+k_{3}-1)!}{(k_{1}-1)!k_{3}%
!}\frac{x_{3}^{k_{3}}(1-x_{3})^{k_{1}-1}}{b_{3}}\exp(a_{1}(1-x_{3})+a_{3}%
x_{3})\\
&  =\frac{1}{n!}\frac{(k_{1}+k_{3}-1)!}{(k_{1}-1)!(k_{3}-1)!}\frac
{x_{3}^{k_{3}-1}(1-x_{3})^{k_{1}-1}}{b_{3}}\exp(a_{1}(1-x_{3})+a_{3}x_{3})
\end{align*}
and analogously for $k_{3}=0$.
\end{proof}

For the case $x\in]b_{3},b_{2}[$ we shall apply the following parametrization%
\[
t\mapsto((1-t)y_{1},(1-y_{1})+t(y_{1}-y_{3}),ty_{3})
\]
for $0\leq y_{1}\leq y_{3}\leq1$ satisfying the relations%
\begin{align*}
b_{2}y_{1}  &  =b_{2}-x\\
(b_{2}-b_{3})y_{3}  &  =b_{2}-x.
\end{align*}
This leads as in the proof of Lemma \ref{originaldensity} to the volume
element%
\[
\frac{1}{\sqrt{b_{2}(b_{2}-b_{3})}}\sqrt{y_{1}y_{3}}dx
\]
under variations of $x$, hence the respective densities $\chi$ satisfy the
following relations: we fix a path $\gamma\in C_{n}$ with characteristic
$\operatorname{char}(\gamma)=(k_{1},k_{2},k_{3})$, $k_{1}+k_{2}+k_{3}=n$ for
$n\geq2$, hence%
\begin{equation}
\chi(k_{1},k_{2},k_{3},x)=\sqrt{\frac{1}{b_{2}(b_{2}-b_{3})}}\frac{1}{n!}%
\int_{0}^{1}f((1-t)y_{1},1-y_{1}+t(y_{1}-y_{3}),ty_{3})\sqrt{y_{1}y_{3}}dt
\label{second interval}%
\end{equation}
in the sense of Gamma-functions.

\begin{remark}
\label{Rem7} Notice that the case $x\in]b_{3},b_{2}[$ is deduced from the case
$x\in]0,b_{2}[$ by the permutation $1\longleftrightarrow2$ is performed. One
replaces then $x_{2}$ by $y_{1}$, $x_{3}$ by $y_{3}$, performs the permutation
for $a_{ij}$, and replaces $b_{2}$ by $b_{2}$and $b_{3}$ by $b_{2}-b_{3}$. All
the necessary relations maintain and the first case in full generality then
implies the second one.
\end{remark}

\section{Combinatorial Sums}

Our next goal is to represent $\psi^{A,B}(x):=\psi(x)$ in the following way.
By Remark \ref{Rem7} it suffices to consider the interval $]0,b_{3}[$.

\begin{proposition}
\label{Le12} Suppose that $b_{2}>b_{3}$. Then, for $x\in]0,b_{3}[$, we have
\begin{align*}
\psi(x)=  &  \sum_{\gamma\in C}\chi(k_{1},k_{2},k_{3},x)\,\operatorname{ord}%
(\gamma)\\
&  =\frac{1}{x}\ \sum_{k\geq1}\ \sum_{m\geq0}\ \sum_{l\geq0,l\equiv
m\bmod2}(1-x_{3})^{k-1}e^{\lambda+\mu}\sum_{L\geq0}\frac{(-\lambda)^{L}}{L!}\\
&  \quad\times\sum_{r=0}^{k-1}\binom{k-1}{r}\left(  \frac{x_{3}-x_{2}}%
{1-x_{3}}\right)  ^{r}\frac{\left(  \frac{2k+m-l}{2}\right)  _{r+L}%
}{k!(k+m+r+L-1)!}\\
&  \quad\times\sum_{0\leq j\leq k,j\equiv m\bmod2}\binom{k}{j}\binom
{\frac{m-j}{2}+k-1}{k-1}\binom{k-j}{\frac{l-j}{2}}\,2^{j}\\
&  \quad\times(a_{12}\sqrt{x_{2}})^{2k-l}\,(a_{13}\sqrt{x_{3}})^{l}%
\,(a_{23}\sqrt{x_{2}x_{3}})^{m}%
\end{align*}

\end{proposition}

%\begin{proof}
%The proof is almost the same as that of Lemma~\ref{Le10}.
%We use Lemma~\ref{Le3} (in particular the second representation of
%Remark~\ref{Rem6}) and Lemma~\ref{Le4}. We also note that
%$k_2+k_3 = n-k$, $k_2 = (2k-l+m)/2$, $k_3 = (l+m)/2$
%\end{proof}

The \textit{proof} of Proposition~\ref{Le12} is just a direct combination of
Remark~\ref{remdiagonal2}, the following Lemma~\ref{Le4} and the
representations $l_{12} = 2k-l$, $k_{2} = (2k-l+m)/2$, and $k_{3} = (l+m)/2$
when $k_{1} = k$, $l_{13}=l$, and $l_{23} = m$ are given.

However, the representation of $\psi(x)$ in Proposition~\ref{Le12} has to be
transformed in a proper way to observe that it is non-negative. For this
purpose we will further introduce the hypergeometric function $F(a,b;c;z)$ and
use certain \textit{hypergeometric identities} in order to simplify the above representation.

%\section{Auxiliary Results from Cominatorics}

\subsection{Counting paths on the triangle}

\begin{lemma}
\label{Le4} The number of paths $\gamma$ in $C$ with $k_{1}(\gamma)=k$,
$l_{13}(\gamma)=l$, $l_{23}(\gamma)=m$ and $l\equiv m\bmod2$ is given through
\begin{equation}
\frac{2k+m}{k}\sum_{0\leq j\leq k,j\equiv m\bmod2}\binom{k}{j}\binom
{\frac{m-j}{2}+k-1}{k-1}\binom{k-j}{\frac{l-j}{2}}\,2^{j}. \label{eqPrep}%
\end{equation}
If $l\ncong m\bmod2$, then the number of paths vanishes.
\end{lemma}

\begin{proof}
From \cite{JacSzp:02} we get that the generating function of
$\operatorname{ord}(\gamma)$ of all paths $\gamma$ with $\gamma_{1} = 1$ is
given by
\begin{align*}
\sum_{\gamma, \gamma_{1} = 1} \operatorname{ord}(\gamma)  &  = \frac{\left|
\begin{array}
[c]{cc}%
1 & -a_{23}\\
-a_{23} & 1
\end{array}
\right|  } {\left|
\begin{array}
[c]{ccc}%
1 & -a_{12} & -a_{13}\\
-a_{12} & 1 & -a_{23}\\
-a_{13} & -a_{23} & 1
\end{array}
\right|  }\\
&  = \frac{1-a_{23}}{1-2a_{12}a_{13}a_{23}- a_{12}^{2}- a_{13}^{2} -a_{23}%
^{2}}.
\end{align*}
Hence, if $P_{1}(k,l,m)$ denotes the number of paths $\gamma$ in $C$ with
$k_{1}(\gamma) = k$, $l_{13}(\gamma) = l$, $l_{23}(\gamma) = m$, and
$\gamma_{1} = 1$ we have
\[
\sum_{k,l,m} P_{1}(k,l,m) x^{k} a_{13}^{l} a_{23}^{m} = \frac{1-a_{23}^{2}%
}{1-a_{23}^{2} -x(2a_{13}a_{23}+ a_{13}^{2}+ 1) }.
\]
From that we immediately get (if $l\equiv m\bmod2$)
\[
P_{1}(k,l,m) = \sum_{0\le j\le k, j\equiv m \bmod2} \binom{k}{j} \binom
{\frac{m-j}2 + k-1}{k-1} \binom{k-j}{\frac{l-j}2}\, 2^{j}.
\]
Finally, if we denote by $P(k,l,m)$ the total number of paths $\gamma$ in $C$
with $k_{1}(\gamma) = k$, $l_{13}(\gamma) = l$, and $l_{23}(\gamma) = m$ then
\[
\frac1k P_{1}(k,l,m) = \frac1n P(k,l,m),
\]
where $n = 2k + m = k_{1}+k_{2}+k_{2}$. This proves (\ref{eqPrep}).
\end{proof}

\subsection{Hypergeometric identities}

The hypergeometric function $F(a,b;c;z)$ is defined (for complex $|z|<1$) by
\[
F(a,b;c;z)=\sum_{n\geq0}\frac{(a)_{n}(b)_{n}}{(c)_{n}n!}z^{n},
\]
where $(x)_{n}=\Gamma(x+n)/\Gamma(x)=x(x+1)\cdots(x+n-1)$ denote the
\textit{rising factorials}. There are lots of identities (see \cite[Chapter
15]{AbrSte:72}) for these kinds of functions. Some of them will be used in the
sequel. For example one has Euler's integral representation
\[
F(a,b;c;z)=\frac{\Gamma(c)}{\Gamma(b)\Gamma(c-b)}\int_{0}^{1}(1-zt)^{-a}%
t^{b-1}(1-t)^{c-b-1}\,dt
\]
if $|z|<1$ and $\Re(c)>\Re(b)>0$. Furthermore, it was already known to Gauss
that
\begin{equation}
F(a,b;c;1)=\frac{\Gamma(c)\Gamma(c-a-b)}{\Gamma(c-a)\Gamma(c-b)}
\label{eqhyp2}%
\end{equation}
if $\Re(c-a-b)>0$.

We start with a lemma, where we use the identity
\begin{equation}
\label{eqH3}F(a,b;c;z) = (1-z)^{-a} F\left(  a,c-b;c,\frac z{z-1}\right)  .
\end{equation}

\begin{lemma}
\label{Le5} Suppose that $j\equiv m \bmod2$. Then
\begin{align*}
&  2^{j} \sum_{l\ge j, l\equiv j \bmod2} \binom{k-j}{\frac{l-j}2} \left(
\frac{2k+m-l}2 \right)  _{r} \, C^{l} D^{2k-l}\\
&  = (C^{2} + D^{2})^{k} v^{j} \sum_{\rho=0}^{r} (-1)^{\rho}\binom{r}{\rho}
\left(  \frac{2k+m-j}2 \right)  _{r-\rho} \frac{(k-j)!}{(k-j-\rho)!} \left(
\frac{C}{2D} v \right)  ^{\rho},
\end{align*}
where $v = 2CD/(C^{2} + D^{2})$.
\end{lemma}

\begin{proof}
We note that the left hand side of the above equation can be represented as
\begin{align*}
&  2^{j} (CD)^{j} d^{2(k-j)} \left(  \frac{2k+m-j}2 \right)  _{r}\\
&  \times F\left(  -(k-j), -\left(  \frac{2k+m-j}2-1\right)  ; -\left(
\frac{2k+m-j}2+r-1\right)  ; -\frac{C^{2}}{D^{2}} \right)
\end{align*}
and the right hand side as
\begin{align*}
&  (C^{2} + D^{2})^{k-j} (2CD)^{j} \left(  \frac{2k+m-j}2 \right)  _{r}\\
&  \times F\left(  -(k-j), -r; -\left(  \frac{2k+m-j}2+r-1\right)  ;
\frac{C^{2}} {C^{2} + D^{2}} \right)  .
\end{align*}
By using (\ref{eqH3}) with
\[
a = -(k-j),\ b= -\left(  \frac{2k+m-j}2-1\right)  ,\ c = -\left(
\frac{2k+m-j}2+r-1\right)
\]
and $z = - {C^{2}}/{D^{2}}$ we directly get a proof of the lemma.
\end{proof}

\subsection{Further Hypergeometric Identities}

In this section we present a proof of rather strange identities that seem to
be new in the context of hypergeometric series.

We set
\[
A_{r}(k;v,\xi) :=\sum_{m\ge0}\sum_{j=0}^{k} {\binom{k}{j}} \frac
{2^{k+r-1}\left(  \frac{m-j+2}2\right)  _{k+r-1}}{(m+1)_{k+2r-1}} \,v^{j}
\,\frac{\xi^{m}}{m!},
\]
where $r$ a is non-negative integer.

\begin{lemma}
\label{Lebasic} We have
\begin{align*}
&  A_{0}(k;v,\xi) = (1+v)^{k}e^{\xi}\\
&  + \int_{0}^{1} \sum_{\ell\ge0} \frac{k!}{\ell!(\ell+1)!(k-2\ell-2)!}
(1+sv)^{k-2\ell-2} \left(  \frac{1-v^{2}}2 \right)  ^{\ell+1} \left(
\frac{1-s^{2}}2 \right)  ^{\ell}\, e^{s\xi}\, ds\\
&  = (1+v)^{k}e^{\xi}\\
&  + \binom k2 (1-v^{2}) \int_{0}^{1} (1+sv)^{k-2} F\left(  -\frac{k-2}2,
-\frac{k-3}2; 2; \frac{(1-v^{2})(1-s^{2})}{(1+sv)^{2}} \right)  e^{s\xi}\, ds
\end{align*}
and
\begin{align*}
&  A_{r}(k;v,\xi)\\
&  = \int_{0}^{1} \sum_{\ell\ge0} \frac{k!}{\ell!(\ell+r-1)!(k-2\ell)!}
(1+sv)^{k-2\ell} \left(  \frac{1-v^{2}}2 \right)  ^{\ell} \left(
\frac{1-s^{2}}2 \right)  ^{\ell+r-1}\, e^{s\xi}\, ds\\
&  = \int_{0}^{1} \frac{(1+sv)^{k}}{(r-1)!}\left(  \frac{1-s^{2}}2 \right)
^{r-1} F\left(  -\frac{k}2, -\frac{k-1}2; r; \frac{(1-v^{2})(1-s^{2}%
)}{(1+sv)^{2}} \right)  e^{s\xi}\, ds,
\end{align*}
where $r$ a is positive integer.
\end{lemma}

\begin{remark}
Note that the right hand sides of these identities are non-negative if
$|v|\le1$. Hence, we have $A_{r}(k;v,\xi)\ge0$.

In fact, we are more interested in sums of the form
\begin{align*}
\tilde A_{r}(k;v,\xi)  &  = \frac12\left(  A_{r}(k;v,\xi) + A_{r}(k;-v,-\xi)
\right) \\
&  = \sum_{m\ge0}\ \sum_{0\le j\le k,\ j\equiv m \bmod2} {\binom{k}{j}}
\frac{2^{k+r-1}\left(  \frac{m-j+2}2\right)  _{k+r-1}}{(m+1)_{k+2r-1}} \,v^{j}
\,\frac{\xi^{m}}{m!}.
\end{align*}
Since $A_{r}(k;v,\xi)\ge0$ and $A_{r}(k;-v,-\xi)\ge0$(for $|v|\le1$) we also
have $\tilde A_{r}(k;v,\xi)\ge0$ and the representations
\begin{align*}
&  \tilde A_{0}(k;v,\xi) = \frac{(1+v)^{k}}2 e^{\xi}+ \frac{(1-v)^{k}}2
e^{-\xi}\\
&  +\binom k2 \frac{1-v^{2}}2 \int_{-1}^{1} (1+sv)^{k-2} F\left(  -\frac
{k-2}2, -\frac{k-3}2; 2; \frac{(1-v^{2})(1-s^{2})}{(1+sv)^{2}} \right)
e^{s\xi}\, ds
\end{align*}
and
\begin{align*}
&  \tilde A_{r}(k;v,\xi)\\
&  = \frac12 \int_{-1}^{1} \frac{(1+sv)^{k}}{(r-1)!}\left(  \frac{1-s^{2}}2
\right)  ^{r-1} F\left(  -\frac{k}2, -\frac{k-1}2; r; \frac{(1-v^{2}%
)(1-s^{2})}{(1+sv)^{2}} \right)  e^{s\xi}\, ds,
\end{align*}
where $r$ a is positive integer.
\end{remark}

\begin{proof}
\!\footnote{This nice proof was pointed out to us by Christian Krattenthaler
and is considerably easier than our first one.} We prove first the case of
positive $r$. Both sides of the identity are power series in $v$ and $\xi$.
Thus, it is sufficient to compare coefficients. The coefficient of $v^{j}%
\xi^{m}/m!$ of the right hand side is given by
\begin{align*}
&  [v^{j}]\int_{0}^{1}\sum_{\ell\geq0}\frac{k!}{\ell!(\ell+r-1)!(k-2\ell
)!}(1+sv)^{k-2\ell}\left(  \frac{1-v^{2}}{2}\right)  ^{\ell}\left(
\frac{1-s^{2}}{2}\right)  ^{\ell+r-1}\,s^{m}\,ds\\
&  =\int_{0}^{1}\sum_{\ell\geq0}\frac{k!}{\ell!(\ell+r-1)!(k-2\ell)!}\\
&  \qquad\qquad\times\sum_{i\geq0}(-1)^{i}\binom{\ell}{i}\frac{1}{2^{\ell}%
}\binom{k-2\ell}{j-2i}s^{j-2i}\left(  \frac{1-s^{2}}{2}\right)  ^{\ell
+r-1}\,s^{m}\,ds
\end{align*}
By applying the substitution $s=\sqrt{t}$, integrating the corresponding Beta
integrals and rewriting the sum over $\ell$ in hypergeometric notation we thus
get
\begin{align*}
&  \int_{0}^{1}\sum_{\ell\geq0}\frac{k!}{\ell!(\ell+r-1)!(k-2\ell)!}\\
&  \qquad\qquad\times\sum_{i\geq0}(-1)^{i}\binom{\ell}{i}\frac{1}{2^{2i+r}%
}\binom{k-2\ell}{j-2i}t^{m/2+j/2-i-1/2}(1-t)^{\ell+r-1}\,dt\\
&  =\sum_{\ell\geq0}\frac{k!}{\ell!(\ell+r-1)!(k-2\ell)!}\\
&  \qquad\qquad\times\sum_{i\geq0}(-1)^{i}\binom{\ell}{i}\frac{1}{2^{2i+r}%
}\binom{k-2\ell}{j-2i}\frac{\Gamma(m/2+j/2-i+1/2)\Gamma(\ell+r)}{\Gamma
(\ell+r+m/2+j/2-i+1/2)}\\
&  =\sum_{i\geq0}\frac{(-1)^{i}(1+i)_{k-i}}{2^{2i+r}(j-2i)!(k-j)!(\frac{1}%
{2}+i+\frac{j}{2}+\frac{m}{2})_{i+r}}\\
&  \qquad\qquad\times F\left(  \frac{j}{2}-\frac{k}{2},\frac{1}{2}+\frac{j}%
{2}-\frac{k}{2};\frac{1}{2}+\frac{j}{2}+\frac{m}{2}+r;1\right)  .
\end{align*}
Next we use formula (\ref{eqhyp2}) and obtain (after rewriting the remaining
sum in hypergeometric notation)
\[
\binom{k}{j}\frac{\Gamma\left(  \frac{1}{2}+\frac{j}{2}+\frac{m}{2}\right)
\Gamma\left(  -\frac{j}{2}+k+\frac{m}{2}+r\right)  }{2^{r}\Gamma\left(
\frac{k}{2}+\frac{m}{2}+r\right)  \Gamma\left(  \frac{1}{2}+\frac{k}{2}%
+\frac{m}{2}+r\right)  }F\left(  -\frac{j}{2},\frac{1}{2}-\frac{j}{2};\frac
{1}{2}-\frac{j}{2}-\frac{m}{2};1\right)  .
\]
In order to avoid difficulties with zero-cancellations we interprete this sum
as a limit, use again formula (\ref{eqhyp2}) and obtain (after some algebra)
\begin{align*}
&  \lim_{\varepsilon\rightarrow0}\binom{k}{j}\frac{\Gamma\left(  \frac{1}%
{2}+\frac{j}{2}+\frac{m}{2}\right)  \Gamma\left(  -\frac{j}{2}+k+\frac{m}%
{2}+r\right)  }{2^{r}\Gamma\left(  \frac{k}{2}+\frac{m}{2}+r\right)
\Gamma\left(  \frac{1}{2}+\frac{k}{2}+\frac{m}{2}+r\right)  }F\left(
-\frac{j}{2},\frac{1}{2}-\frac{j}{2};\frac{1}{2}-\frac{j}{2}-\frac{m}%
{2}+\varepsilon;1\right) \\
&  =\lim_{\varepsilon\rightarrow0}\binom{k}{j}\frac{2^{k+r+2\varepsilon
-2}\Gamma\left(  -\frac{j}{2}+k+r+\frac{m}{2}\right)  \Gamma\left(
m-2\varepsilon+1\right)  \sin(\pi(2\varepsilon-m))}{(k+m+2r)!\Gamma\left(
1-\frac{j}{2}+\frac{m}{2}-\varepsilon\right)  \sin\left(  \pi\left(  \frac
{1}{2}-\frac{j}{2}-\frac{m}{2}+\varepsilon\right)  \right)  \sin\left(
\pi\left(  \frac{j}{2}-\frac{m}{2}+\varepsilon\right)  \right)  }.
\end{align*}
Now note that the limit of the $\sin$-terms is always $2$. Hence, we finally
obtain
\[
\frac{2^{k+r-1}\left(  \frac{m-j+2}{2}\right)  _{k+r-1}}{(m+1)_{k+2r-1}}%
\]
as proposed.

The proof for the case $r=0$ runs along similar lines. The only difference is
the \textit{singular term} $\binom kj$ in front. However, after integrating
the Beta integrals we can rewrite the corresponding sum as
\begin{align*}
&  \binom kj + \sum_{\ell\ge0} \frac{k!}{\ell!(\ell+1)!(k-2\ell-2)!}\\
&  \qquad\qquad\times\sum_{i\ge0} (-1)^{i} \binom{\ell+1} k \frac1{2^{2\ell
+2}} \binom{k-2\ell-2}{j-2i} \frac{\Gamma(m/2+j/2-i+1/2)\Gamma(\ell+1)}
{\Gamma(\ell+m/2+j/2-i+3/2)}\\
&  = \sum_{\ell\ge-1} \frac{k!}{\ell!(\ell+1)!(k-2\ell-2)!}\\
&  \qquad\qquad\times\sum_{i\ge0} (-1)^{i} \binom{\ell+1} k \frac1{2^{2\ell
+2}} \binom{k-2\ell-2}{j-2i} \frac{\Gamma(m/2+j/2-i+1/2)\Gamma(\ell+1)}
{\Gamma(\ell+m/2+j/2-i+3/2)}\\
&  =\sum_{i\ge0} \frac{(-1)^{i} (1+j-2i)_{k-j+2i}} {2^{2i}(k-j)!i!(\frac12
-i+\frac j2 + \frac m2)_{i}} F\left(  \frac j2 - \frac k2, \frac12 + \frac j2
- \frac k2; \frac12 + \frac j2 + \frac m2; 1 \right)
\end{align*}
and proceed as above.
\end{proof}

\begin{lemma}
\label{Lereps} Set
\begin{equation}
T(k,r,\rho;v,\xi): = \sum_{m\ge0}\sum_{j=0}^{k} {\binom{k}{j}} \frac
{2^{k+r-\rho-1}\left(  \frac{m-j+2}2\right)  _{k+r-\rho-1}}{(m+1)_{k+r-1}}
\,\frac{(k-j)!}{(k-j-\rho)!}\,v^{j} \,\frac{\xi^{m}}{m!}%
\end{equation}
and
\begin{equation}
\label{eqLeresp2}S(k,r,\rho;v,\xi): = \sum_{\tau=0}^{r-\rho} (-1)^{r-\rho
-\tau} \binom{r-\rho}\tau T(k,r,\rho+\tau;v,\xi).
\end{equation}
Then
\begin{equation}
\label{eqLeresp3}T(k,r,\rho;v,\xi) = \sum_{\tau=0}^{r-\rho} \binom{r-\rho}\tau
S(k,r,\rho+\tau;v,\xi)
\end{equation}
and
\begin{equation}
\label{eqLeresp4}S(k,r,\rho;v,\xi) = \frac{k!}{(k-\rho)!} \sum_{a\ge0}
\binom{r-\rho}{2a} \frac{(2a)!}{2^{a} a!} A_{a+\rho}(k-\rho;v,\xi).
\end{equation}

\end{lemma}

In particular we have $S(k,r,\rho;v,\xi)\ge0$ and $T(k,r,\rho;v,\xi)\ge0$ if
$|v|\le1$.

\begin{remark}
If we set $\tilde T(k,r,\rho;v,\xi) = \frac12 (T(k,r,\rho;v,\xi) +
T(k,r,\rho;-v,-\xi))$ and $\tilde S(k,r,\rho;v,\xi) = \frac12 (S(k,r,\rho
;v,\xi) + S(k,r,\rho;-v,-\xi))$ then we have (of course) corresponding
representations in terms of $\tilde A_{r}(k;v,\xi)$ and also $\tilde
S(k,r,\rho;v,\xi)\ge0$ and $\tilde T(k,r,\rho;v,\xi)\ge0$ if $|v|\le1$.
\end{remark}

\begin{proof}
First note that (\ref{eqLeresp2}) and (\ref{eqLeresp3}) are equivalent. Thus,
it remains to prove (\ref{eqLeresp4}) or equivalently
\begin{align*}
&  T(k,r,\rho;v,\xi)\\
&  = \sum_{\tau=0}^{r-\rho} \binom{r-\rho}\tau\frac{k!}{(k-\rho-\tau)!}
\sum_{a\ge0} \binom{r-\rho-\tau}{2a} \frac{(2a)!}{2^{a} a!} A_{a+\rho+\tau
}(k-\rho-\tau;v,\xi).
\end{align*}
By expanding both sides with respect to $v^{j} \xi^{m}/m!$ this identitiy is
equivalent to
\begin{align*}
&  {\binom{k}{j}} \frac{2^{k+r-\rho-1}\left(  \frac{m-j+2}2\right)
_{k+r-\rho-1}}{(m+1)_{k+r-1}} \,\frac{(k-j)!}{(k-j-\rho)!}\\
&  =\sum_{\tau,a\ge0} \binom{r-\rho}\tau\frac{k!}{(k-\rho-\tau)!}
\binom{r-\rho-\tau}{2a} \frac{(2a)!}{2^{a} a!} {\binom{k-\rho-\tau}{j}}
\frac{2^{k+a-1}\left(  \frac{m-j+2}2\right)  _{k+a-1}}{(m+1)_{k+2a+\rho
+\tau-1}}%
\end{align*}
By rewriting the sum over $\tau$ of the right hand side in hypergeometric
notation and by using (\ref{eqhyp2}) we get
\begin{align*}
&  \sum_{a\ge0} \frac{(r-\rho)! k! 2^{k-1}\left(  \frac{m-j+2}2\right)
_{k+a-1}} {j!(k-\rho-j)!(r-\rho-2a)!a! (m+1)_{k+2a+\rho-1}}\\
&  \qquad\qquad\times F\left(  -(r-\rho-2a),-(k-\rho-j); m+k+2a+\rho;1\right)
\\
&  =\sum_{a\ge0} \frac{(r-\rho)! k! 2^{k-1}\left(  \frac{m-j+2}2\right)
_{k+a-1}} {j!(k-\rho-j)!(r-\rho-2a)!a! (m+1)_{k+2a+\rho-1}}\\
&  \qquad\qquad\times\frac{\Gamma(m+k+2a+\rho)\Gamma(m+2k+r-\rho-j)}
{\Gamma(m+k+r)\Gamma(m+2k+2a-j)}%
\end{align*}
Next this sum can be also written in hypergeometric notation. Further, a
second use of (\ref{eqhyp2}) and some simplifications (using the duplication
formula of the Gamma functions) yield
\begin{align*}
&  \frac{k!2^{k-1}\left(  \frac{m-j+2}2\right)  _{k-1}m! \Gamma(m+2k+r-\rho
-j)} {j!(k-\rho-j)! \Gamma(m+k+r)\Gamma(m+2k-j)}\\
&  \qquad\qquad\times F\left(  -\frac{r-\rho}2,-\frac{r-\rho-1}2;
\frac{m+2k-j+1}2;1 \right) \\
&  = \binom kj \frac{(k-j)!2^{k-1}\left(  \frac{m-j+2}2\right)  _{k-1}
\Gamma(m+2k+r-\rho-j)}{(k-\rho-j)!(m+1)_{k+r-1}\Gamma(m+2k-j) }\\
&  \qquad\qquad\times\frac{\Gamma\left(  k+\frac m2 - \frac j2 +\frac12
\right)  \Gamma\left(  k + \frac m2 - \frac j2 + r- \rho-1 \right)  }
{\Gamma\left(  k + \frac m2 - \frac j2 + \frac r2 - \frac\rho2 - \frac12
\right)  \Gamma\left(  k + \frac m2 - \frac j2 + \frac r2 - \frac\rho2 - 1
\right)  }\\
&  = {\binom{k}{j}} \frac{2^{k+r-\rho-1}\left(  \frac{m-j+2}2\right)
_{k+r-\rho-1}}{(m+1)_{k+r-1}} \,\frac{(k-j)!}{(k-j-\rho)!}%
\end{align*}
as proposed.
\end{proof}

\section{Proof of Theorem~\ref{TH}}

First we use the results of the previous section to obtain another
representation for $\psi(x)$.

\begin{lemma}
\label{Le13} Suppose that $b_{2} > b_{3}$ and set
\begin{align*}
A_{12}  &  = a_{12}\sqrt{x_{2}} = a_{12}\sqrt{\frac x{b_{2}}},\\
A_{13}  &  = a_{13}\sqrt{x_{3}} = a_{13}\sqrt{\frac x{b_{3}}},\\
v  &  = \frac{2A_{12}A_{13}}{A_{12}^{2}+ A_{13}^{2}},\\
\xi &  = a_{23}\sqrt{x_{2}x_{3}},\\
w_{1}  &  = \frac{(1-x_{3})}2 (A_{12}^{2}+ A_{13}^{2}),\\
w_{2}  &  = \frac{x_{3}-x_{2}}{2(1-x_{3})}\\
\omega &  = 1- \frac{A_{13}}{A_{12}} v = \frac{A_{12}^{2} - A_{13}^{2}}%
{A_{12}^{2}+ A_{13}^{2}}.
\end{align*}
Then for $x\in]0,b_{3}[$ we have
\begin{align*}
\psi(x)  &  = \frac{2e^{\lambda+\mu}}{x(1-x_{3})} \sum_{k\ge1} \frac{w_{1}%
^{k}}{k!(k-1)!} \sum_{r=0}^{k-1} \binom{k-1}r \, w_{2}^{r} \sum_{L\ge0}
\frac{(-\lambda)^{L}}{L!}\\
&  \quad\times\sum_{\rho=0}^{r+L} \binom{r+L}{\rho}\tilde S(k,r+L,\rho;v,\xi)
\, \omega^{\rho}.
\end{align*}

\end{lemma}

\begin{proof}
With help of Proposition~\ref{Le12} and Lemma~\ref{Le5} we get
\begin{align*}
\psi(x)  &  = \frac{2e^{\lambda+\mu}} {x(1-x_{3})} \sum_{k\ge1} \frac
{w_{1}^{k}}{k!} \sum_{r=0}^{k-1} \binom{k-1}r \ w_{2}^{r} \sum_{L\ge0}
\frac{(-\lambda)^{L}}{L!}\\
&  \quad\times\sum_{m\ge0}\ \sum_{0\le j\le k, j\equiv m \bmod2} \binom kj
\binom{\frac{m-j}2 + k-1}{k-1}\\
&  \quad\times\sum_{\rho=0}^{r+L} (-1)^{\rho}\binom{r+L}{\rho} 2^{k+r+L-\rho
-1} \left(  \frac{2k+m-j}2 \right)  _{r+L-\rho} \frac{(k-j)!}{(k-j-\rho)!}\\
&  \quad\times v^{j}\, \left(  \frac{A_{13}}{A_{12}} v \right)  ^{\rho}\,
\frac{\xi^{m}}{(m+k+r-1)!}\\
&  = \frac{2e^{\lambda+\mu}} {x(1-x_{3})} \sum_{k\ge1} \frac{w_{1}^{k}%
}{k!(k-1)!} \sum_{r=0}^{k-1} \binom{k-1}r \ w_{2}^{r} \sum_{L\ge0}
\frac{(-\lambda)^{L}}{L!}\\
&  \quad\times\sum_{\rho=0}^{r+L} (-1)^{\rho}\binom{r+L}{\rho} \left(
\frac{A_{13}}{A_{12}} v \right)  ^{\rho}\\
&  \quad\times\sum_{m\ge0}\ \sum_{0\le j\le k, j\equiv m \bmod2} \binom kj
\frac{2^{k+r+L-\rho-1}\left(  \frac{m-j+2}2 \right)  _{k+r+L-\rho-1}}
{(m+1)_{k+r+L-1}}\frac{(k-j)!}{(k-j-\rho)!}\ v^{j} \frac{\xi^{m}}{m!}\\
&  = \frac{2e^{\lambda+\mu}}{x(1-x_{3})} \sum_{k\ge1} \frac{w_{1}^{k}%
}{k!(k-1)!} \sum_{r=0}^{k-1} \binom{k-1}r \ w_{2}^{r} \sum_{L\ge0}
\frac{(-\lambda)^{L}}{L!}\\
&  \quad\times\sum_{\rho=0}^{r+L} (-1)^{\rho}\binom{r+L}{\rho} \left(
\frac{A_{13}}{A_{12}} v \right)  ^{\rho} \tilde T(k,r+L,\rho;v,\xi)
\end{align*}
Finally by using (\ref{eqLeresp3}) we directly derive the proposed representation.
\end{proof}

Note that $|v|\leq1$. Thus this lemma shows that $\psi(x)\geq0$ if $\omega
\geq0$ and $\lambda\leq0$ or equivalently $|A_{12}|\geq|A_{13}|$ and
$a_{1}(b_{2}-b_{3})+a_{2}b_{3}-a_{3}b_{2}\geq0$. This is satisfied by
assumptions of Theorem \ref{TH}. Hence we have proved Theorem~\ref{TH} for
$x\in]0,b_{3}[$. The case $x\in]b_{3},b_{2}[$ is done by exchanging indices
$1$ and $2$ and $b_{3}$ by $b_{2}-b_{3}$ (compare with Remark \ref{Rem7}).

\section{Appendix 1: the Feynman-Kac formula}

We shall work with a continuous-time Markov process on the state space
$S=\mathbb{C}\times\{1,\dots,d\}$ associated to the off-diagonal elements of a
$d\times d$ matrix $A\in M_{d}(\mathbb{C})$ with complex entries. The non-zero
diagonal elements of $A$ and the \textquotedblright
potential\textquotedblright\ $B$ will appear in exponential functionals of the
process. We denote $\mathbb{C}$-valued functions on the state space $S$ by
$f$. Given a matrix $A\in M_{d}(\mathbb{C})$ with zero diagonal, we associate
a generator $\mathcal{A}$ of a pure-jump type process on $S$, namely%
\[
\mathcal{A}f(\zeta,i)=\sum_{\substack{j=1\\j\neq i}}^{d}(f(\zeta
a_{ij},j)-f(\zeta,i)).
\]
The resulting $S$-valued Markov process is denoted by $Y_{t}^{(\zeta
,i)}:=(Z_{t}^{(\zeta,i)},X_{t}^{(\zeta,i)})$ with initial value $(\zeta,i)\in
S$ and has the following properties.

\begin{lemma}
The projection of $(Y_{t}^{(\zeta,i)})_{t\geq0}=:(Z_{t}^{(\zeta,i)}%
,X_{t}^{(\zeta,i)})_{t\geq0}$ onto the second component will be denoted by
$(X_{t}^{i})$, since it does not depend on $\zeta\in\mathbb{C}$, and equals in
distribution the $\{1,\dots,d\}$-valued Markov process associated to the
matrix $G$ with entries $g_{ij}=1$ for $j\neq i$ and $g_{ii}=1-d$ for
$i,j=1,\dots,d$.
\end{lemma}

\begin{proof}
Take a function $f$ on $S$, which does not depend on the first component
$\zeta$, then we have%
\begin{align*}
\mathcal{A}f(i)  &  =\sum_{j=1}^{d}(f(j)-f(i))\\
&  =(Gf)_{i},
\end{align*}
where we can identify $f$ with a vector in $\mathbb{C}^{d}$.
\end{proof}

\begin{lemma}
Let $f^{r}:\mathbb{C}\times\{1,\dots,d\}\rightarrow\mathbb{C}$ be the function%
\[
f^{r}(\zeta,i):=\zeta r_{i}%
\]
for $(\zeta,i)\in S$ and $r\in\mathbb{C}^{d}$ (which may be identified with a
linear map from $\mathbb{C}^{d}$ to $C(S,\mathbb{C})$), then%
\[
\mathcal{A}f^{r}=f^{(A-(d-1)1_{d})r}.
\]

\end{lemma}

\begin{proof}
Direct calculation of the generator $\mathcal{A}$ on functions $f^{r}$.
\end{proof}

Furthermore given a real valued function $V^{b}$ on $S$ of the form%
\[
V^{b}(\zeta,i)=b_{i}%
\]
for $(\zeta,i)\in S$ (playing the role of a \textquotedblright
potential\textquotedblright) and $b\in\mathbb{C}^{d}$, then we can define the
multiplication operator (denoted again by $V^{b}$) and we obtain%
\[
V^{b}f^{r}=f^{Br},
\]
where $B$ denotes the diagonal matrix with entries $b_{1},\dots,b_{d}%
\in\mathbb{C}$.

The main assertion of this section is the Feynman-Kac formula for the family
of Markov process $(Y_{t}^{(\zeta,i)})_{t\geq0}$ for $(\zeta,i)\in S$. Since
we do not want to go into the theory of Feller semigroups and maximal domains,
we formulate the Feynman-Kac Theorem in a special case, i.e. on a finite
dimensional domain of definition.

\begin{theorem}
Let $A$ be any $d\times d$ matrix with entries in $\mathbb{C}$ and diagonal
elements $a_{1},\dots,a_{d}$, let $V^{b}$ be the multiplication operator for
$b\in\mathbb{C}^{d}$ as defined above, then for $z\in\mathbb{C}$ the function,%
\[
u_{t}^{r}(\zeta,i)=E\left(  \exp(\int_{0}^{t}a(X_{s}^{i})ds-z\int_{0}%
^{t}b(X_{s}^{i})ds)f^{r}(Y_{t}^{(\zeta,i)})\right)
\]
solves the following differential equation on $C(S,\mathbb{C})$ for initial
value $f^{r}$, $r\in\mathbb{C}^{d}$, and all $t\geq0,$%
\begin{align}
\frac{\partial}{\partial t}u_{t}  &  =\mathcal{A}u_{t}+V^{a}u_{t}-zV^{b}%
u_{t}\label{diffequ}\\
u_{0}  &  =f.
\end{align}
On the other hand, for initial value $f=f^{r}$, the solution to the
differential equation (\ref{diffequ}) is given by%
\[
t\mapsto e^{-t(d-1)}f^{\exp(t(A-zB))r},
\]
hence%
\[
E\left(  \exp(\int_{0}^{t}a(X_{s}^{i})ds-z\int_{0}^{t}b(X_{s}^{i}%
)ds)f^{r}(Y_{t}^{(1,i)})\right)  =e^{-t(d-1)}(\exp(t(A-zB))r)_{i}%
\]
for all $z\in\mathbb{C}$.\label{fk2}
\end{theorem}

\begin{proof}
The proof is done by the Markov property for the process $(Y_{t}^{(\zeta
,i)})_{t\geq0}$ with $(\zeta,i)\in S$: first we show that the following
semigroup property holds:%
\[
u_{t_{1}+t_{2}}^{r}(\zeta,i)=E(\exp(\int_{0}^{t_{2}}a(X_{s}^{i})ds-z\int
_{0}^{t_{2}}b(X_{s}^{i})ds)u_{t_{1}}^{r}(Y_{t_{2}}^{(\zeta,i)}))
\]
for $t_{1},t_{2}\geq0$. Indeed, the right hand side can be written by the
Markov property%
\begin{gather*}
E(\exp(\int_{0}^{t_{2}}a(X_{s}^{i})ds-z\int_{0}^{t_{2}}b(X_{s}^{i}%
)ds)u_{t_{1}}^{r}(Y_{t_{2}}^{(\zeta,i)}))\\
=E\left(  \exp(\int_{0}^{t_{2}}a(X_{s}^{i})ds-z\int_{0}^{t_{2}}b(X_{s}%
^{i})ds)E(\exp(\int_{0}^{t_{1}}a(X_{s}^{i})ds-\right. \\
\left.  -z\int_{0}^{t_{1}}b(X_{s}^{\widetilde{i}})ds)f^{r}(Y_{t_{1}%
}^{(\widetilde{\zeta},\widetilde{i})}))|_{(\widetilde{\zeta},\widetilde
{i})=Y_{t_{2}}^{(\zeta,i)}}\right) \\
=E\left(  \exp(\int_{0}^{t_{2}}a(X_{s}^{i})ds-z\int_{0}^{t_{2}}b(X_{s}%
^{i})ds)E(\exp(\int_{t_{1}}^{t_{1}+t_{2}}a(X_{s}^{i})ds-\right. \\
\left.  -z\int_{t_{1}}^{t_{1}+t_{2}}b(X_{s}^{i})ds)f^{r}(Y_{t_{1}+t_{2}%
}^{(\zeta,i)})|\mathcal{F}_{t_{1}})\right) \\
=E(\exp(\int_{0}^{t_{1}+t_{2}}a(X_{s}^{i})ds-z\int_{0}^{t_{1}+t_{2}}%
b(X_{s}^{i})ds))f^{r}(Y_{t_{1}+t_{2}}^{(\zeta,i)}))\\
=f_{t_{1}+t_{2}}^{r}(\zeta,i)
\end{gather*}
for all $t_{1},t_{2}\geq0$, $(\zeta,i)\in S$ and $r\in\mathbb{C}^{d}$.
Furthermore $r\mapsto u_{t}^{r}(\zeta,i)$ is obviously linear in $r$ for fixed
$(\zeta,i)\in S$ and $t\geq0$. Hence it is sufficient to calculate the
derivative at $t=0$. This leads to%
\begin{align*}
\frac{d}{dt}|_{t=0}u_{t}^{r}(\zeta,i)  &  =a_{i}r_{i}\zeta-zb_{i}r_{i}%
\zeta+\mathcal{A}u_{0}^{r}(\zeta,i)\\
&  =(\mathcal{A}+V^{a}-V^{zb})f^{r}(\zeta,i)\\
&  =f^{(Ar-(d-1)r-zBr)}(\zeta,i).
\end{align*}
Hence $(u_{t}^{r})_{t\geq0,r\in\mathbb{C}^{d}}$ defines a strongly continuous
semiflow on the space $\{f^{r},r\in\mathbb{C}^{d}\}$ with generator
$\mathcal{A}+V^{a}-zV^{b}$, which a fortiori coincides with the semiflow
$t\mapsto e^{-t(d-1)}f^{\exp(t(A-zB))r}$.
\end{proof}

\section{Appendix 2: the first conjecture}

We provide a counter-example for the original version of the BMV-conjecture as
stated in \cite{BesMouVil:75}. We work with the notations of Section 3. Given
the matrices%
\[
A=\left(
\begin{array}
[c]{ccc}%
0 & \frac{\epsilon^{2}}{2} & \epsilon\\
\frac{\epsilon^{2}}{2} & 0 & -\epsilon\\
\epsilon & -\epsilon & 0
\end{array}
\right)
\]
and%
\[
B=\left(
\begin{array}
[c]{ccc}%
0 & 0 & 0\\
0 & 1 & 0\\
0 & 0 & 0
\end{array}
\right)  ,
\]
we define the signed measure $\mu_{1}^{A,B}(dx)$ as inverse Laplace transform
of $z\mapsto\left\langle e_{1},\exp(A-zB)e_{1}\right\rangle $. In the
notations of Section 3 we have $b_{1}=b_{3}=0$ and $b_{2}=1$. We calculate the
sign of the absolutely continuous part of $\mu_{1}^{A,B}$ asymptotically in
$\epsilon$ and show that we obtain a negative sign for small $\epsilon$. We
apply the formulas in the sense of Remark \ref{Rem7}. We notice that
$y_{1}=y_{3}=1-x$ for $0\leq x\leq1$, which leads to the formula%
\[
\phi(k_{1},k_{2},k_{3},x)=\frac{1-x}{(n-1)!}\int_{0}^{1}%
f((1-t)(1-x),x,t(1-x))\frac{(1-t)(1-x)}{k_{1}}dt,
\]
for trajectories $\gamma$ and $\gamma_{1}=1$, hence by Corollary \ref{fk3}
\[
\psi(x)=\sum_{\substack{\gamma\in C\\n\geq2,\gamma_{1}=1}}\phi(k_{1}%
,k_{2},k_{3},x)a_{12}^{l_{12}}a_{13}^{l_{13}}a_{23}^{l_{23}}%
\]
We only have to calculate the following cases up to order $\epsilon^{4}$,
where $\#$ denotes the number of possible paths $\gamma$ with given $l_{ij}$,
where we apply Lemma \ref{Le4}, hence $k_{1}=\frac{l_{12}+l_{13}}{2}\geq1$ and
$2l_{12}+l_{13}+l_{23}\leq4$. This leads to the following table,%
\[%
\begin{tabular}
[c]{|c|c|c|c|}\hline
$l_{12}$ & $l_{13}$ & $l_{23}$ & $\#$\\\hline
$2$ & $0$ & $0$ & $P_{1}(1,0,0)=1$\\\hline
$1$ & $1$ & $1$ & $P_{1}(1,1,1)=2$\\\hline
$0$ & $2$ & $2$ & $P_{1}(1,2,2)=1$\\\hline
$0$ & $4$ & $0$ & $P_{1}(2,4,0)=1$\\\hline
\end{tabular}
\
\]
associated to the paths $121$; $1321$, $1231$; $13231$; $13131$. The fourth
path leads to vanishing $\phi$ since the vertex $2$ is not visited (see Lemma
\ref{originaldensity}, 3. in the appropriate translation as in Remark
\ref{Rem7}). Hence we obtain up to order $\epsilon^{4}$ the following density
for the absolutely continuous part%
\[
\psi_{1}^{A,B}(x)=(1-x)(\frac{\epsilon^{2}}{2})^{2}-2(1-x)^{2}\epsilon
^{2}\frac{\epsilon^{2}}{2}\frac{1}{2}+(1-x)^{3}\epsilon^{4}\frac{1}%
{6}+\mathcal{O}(\epsilon^{5}),
\]
consequently%
\[
\frac{12\psi_{1}^{A,B}(x)}{\epsilon^{4}}=3(1-x)-6(1-x)^{2}+2(1-x)^{3}%
+\mathcal{O}(\epsilon)
\]
where we obtain a negative sign if $x$ is near $0$.

We could also calculate in the original stochastic language by analysing jump
densities and calculating with conditional expectations, which is instructive
in the given example. We apply the formula of Corollary \ref{fk3} and the
definitions of Section 2,%
\begin{gather*}
e^{-2}(\exp(A-zB)e_{1})_{1}=E\left(  \exp(\int_{0}^{1}a(X_{s})ds-z\int_{0}%
^{1}b(X_{s})ds)f^{r}(Y_{1})|X_{0}=1\right)  \\
=E\left(  \exp(-z\int_{0}^{1}1_{\{X_{s}=2\}}ds)f^{r}(Y_{1})|X_{0}=1\right)  \\
=\sum_{\substack{\gamma\in C\\\gamma_{1}=1\\2l_{12}+l_{13}+l_{23}\leq
4}}E_{P_{\gamma}}(\exp(-z\int_{0}^{1}1_{\{X_{s}=2\}}ds)|\gamma_{1}%
=1)P(p_{n}^{-1}(\gamma)|\gamma_{1}=1)a_{12}^{l_{12}}a_{13}^{l_{13}}%
a_{23}^{l_{23}}+\mathcal{O}(\epsilon^{5}),
\end{gather*}
where $(X_{s})_{0\leq s\leq1}$ is the standard Poisson process with intensity
$(d-1)$ on the state space $\{1,2,3\}$. Now we can evaluate the sum directly:

\begin{itemize}
\item the probability for a path configuration $\gamma$ with $n$ jumps is
$e^{-2}\frac{1}{n!}$, see formula \ref{prob-n-jumps} conditioned on
$\gamma_{1}=1$. The duration under the condition that the path configuration
is $\gamma$ is uniformly distributed on the simplex $\Delta_{n}$ (see Section
2 for the precise statement). Rescaling due to the factor $e^{-2}$ on both
sides leads to the precise formulas.

\item the loop $1$ leads to the singular part of $\mu_{1}^{A,B}$.

\item the loop $121$ yields the first non-trivial term in $\psi_{1}^{A,B}$.
This non-trivial contribution is the distribution of the total time in state
$2$ under the condition that the trajectory is $121$. The density of the
projection from $\Delta_{2}$, if we fix the uniform distribution on the
$2$-simplex,%
\begin{align*}
P(T_{2}-T_{1} &  \leq x)=\int_{\Delta_{2},t_{2}\leq x}d\lambda_{2}(t_{1}%
,t_{2},t_{3})\\
&  =2\int_{0}^{x}\int_{0}^{1-t_{2}}dt_{1}dt_{2}=2(x-\frac{x^{2}}{2}).
\end{align*}
The rescaled probability of $121$ is $\frac{1}{2!}$. The density of the second
projection is $2(1-x)$, hence the first contribution $(1-x)a_{12}^{2}$.

\item the loops $1321$, $1231$ yield the second non-trivial contribution, each
of it coming from the density of one projection from the $3$-simplex
$\Delta_{3}$, which yields in the first case%
\begin{align*}
P(T_{3}-T_{2} &  \leq x)=\int_{\Delta_{3},t_{2}\leq x}d\lambda_{3}(t_{1}%
,t_{2},t_{3},t_{4})\\
&  =6\int_{0}^{x}\int_{0}^{1-t_{2}}\int_{0}^{1-t_{3}-t_{2}}dt_{1}dt_{3}%
dt_{2}\\
&  =1-(1-x)^{3},
\end{align*}
and in the equal second case $P(T_{2}-T_{1}\leq x)=1-(1-x)^{3}$. The rescaled
probability of each path $1321$, $1231$ is $\frac{1}{3!}$, hence for the two
paths we obtain the contribution $(1-x)^{2}a_{12}a_{13}a_{23}$.

\item the loop $13231$ yields similarly via the density of the second
projection the function $4(1-x)^{3}$ with rescaled probability $\frac{1}{4!}$
the contribution $\frac{1}{3!}(1-x)^{3}a_{13}^{2}a_{12}$.
\end{itemize}


\begin{thebibliography}{99}                                                                                               %


\bibitem {AbrSte:72}\emph{Handbook of mathematical functions with formulas,
graphs, and mathematical tables,} edited by Milton Abramowitz and Irene A.
Stegun. Dover Publications, Inc., New York, 1972.

\bibitem {BesMouVil:75}Bessis, D., Moussa, P. and Villani, M., \emph{Monotonic
converging variational approximations to the functional integrals in quantum
statistical mechanics}, J. Math. Phys. 16, pp. 2318--2325, 1975.

\bibitem {FanPet:01}Fannes, M., Petz, D., \emph{Perturbation of Wigner
matrices and a conjecture}, Proc. Amer. Math. Soc. 131, no. 7, 1981--1988, 2003.

\bibitem {FanWer:00}Fannes, M., Werner, R., \emph{On some inequalities for the
trace of exponentials}, working paper, 2000.

\bibitem {Fel:66}Feller, W., \emph{An introduction to probability theory and
its applications}, Vol. II, New York-London-Sydney: John Wiley and Sons, Inc.
XVIII, 1966.

\bibitem {JacSzp:02}Jacquet, P.\ and Szpankowski, W,. \emph{A combinatorial
problem arising in information theory: precise minimax redundancy for Markov
sources}. Mathematics and computer science, II (Versailles, 2002), 311--328,
Trends Math., Birkh\"{a}user, Basel, 2002.

\bibitem {KriMic:97}Kriegl, Andreas, Michor, Peter W., \emph{The convenient
Setting of Global Analysis}, Mathematical Surveys and Monographs, vol. 53, 1997.

\bibitem {LieSei:03}Lieb, E. and Seiringer, R., \emph{Equivalent forms of the
Bessis-Moussa-Villani conjecture}, J. Stat. Phys. \textbf{115}, 185-190
(2004). arXiv math-ph/0210027.

\bibitem {Var02}Varadhan, S.R. Srinivasa, PDEs in Finance,
"www.math.nyu.edu/faculty/varadhan/pdefin.html", section 5, 2002.

\bibitem {Wer03}Werner, R., \emph{Open problem list},
"www.imaph.tu-bs.de/home/werner/problems.html", 2003.
\end{thebibliography}
\end{document}